\documentclass[twocolumn]{openjournal}

\usepackage{lipsum}

\usepackage{textcomp}
\usepackage{xcolor}
\usepackage{textgreek}
\usepackage[utf8]{inputenc}
\usepackage[english]{babel}

\usepackage{hyperref}
\hypersetup{
    unicode, 
    colorlinks=true,
    linkcolor=linkcolor,
    citecolor=linkcolor,
    filecolor=linkcolor,
    urlcolor=linkcolor,
}
\usepackage{color,colortbl}
\definecolor{linkcolor}{rgb}{0.0,0.3,0.5}
\usepackage{tensind}
\tensordelimiter{?}
\DeclareGraphicsExtensions{.bmp,.png,.jpg,.pdf}
\usepackage{verbatim}
\usepackage[normalem]{ulem}
\usepackage{orcidlink}
\usepackage{soul}
\newcommand{\Ie}{\ensuremath{I_{\mathrm{E}}}}
\newcommand{\Ye}{\ensuremath{Y_{\mathrm{E}}}}
\newcommand{\Je}{\ensuremath{J_{\mathrm{E}}}}
\newcommand{\He}{\ensuremath{H_{\mathrm{E}}}}
\newcommand{\zphot}{\ensuremath{z_{\mathrm{phot}}}}
\newcommand{\zspec}{\ensuremath{z_{\rm spec}}}

\defcitealias{Stevens25}{GS25}
\defcitealias{Tucci25}{MT25}
\defcitealias{Siudek2025}{MS25}
\defcitealias{Roster2025}{WR25}

\newcommand{\orcidauthor}[3]{\author{\href{http://orcid.org/#1}{#2$^{#3}$}}}

\begin{document}
\title{DeepDISC-Euclid: Source Classification and Photometric Redshifts in Euclid Deep Field North With a Pixel-Level Deep Learning Approach\vspace{-4em}}
\orcidauthor{0009-0001-5163-5781}{Yuanzhe Jiang}{1,*}
\orcidauthor{0000-0003-1659-7035}{Yue Shen}{1,2,*}
\orcidauthor{0009-0005-7923-054X}{Grant Merz}{1}
\orcidauthor{0009-0000-5381-7039}{Shurui Lin}{1}
\orcidauthor{0000-0003-0049-5210}{Xin Liu}{1,2,3,*}
\orcidauthor{0000-0003-0230-6436}{Zhiwei Pan}{1}
\orcidauthor{0000-0001-5105-2837}{Mingyang Zhuang}{1}
\orcidauthor{0000-0002-9149-6528}{William Roster}{4}
\orcidauthor{0000-0001-7116-9303}{Mara Salvato}{4}
\orcidauthor{}{Malgorzata Siudek}{5}
\orcidauthor{0000-0002-8885-4443}{Grant Stevens}{6}

\thanks{$^*$ Corresponding Authors: Yuanzhe Jiang (\href{mailto:yuanzhe@illinois.edu}{yuanzhe@illinois.edu}), Yue Shen (\href{mailto:shenyue@illinois.edu}{shenyue@illinois.edu}), Xin Liu (\href{mailto:xinliuxl@illinois.edu}{xinliuxl@illinois.edu}).}

\affiliation{$^{1}$Department of Astronomy, University of Illinois at Urbana-Champaign, 1002 West Green Street, Urbana, IL 61801, USA}
\affiliation{$^{2}$National Center for Supercomputing Applications, University of Illinois at Urbana-Champaign, 1205 West Clark Street, Urbana, IL 61801, USA}
\affiliation{$^{3}$Center for Artificial Intelligence Innovation, University of Illinois at Urbana-Champaign, 1205 West Clark Street, Urbana, IL 61801, USA}
\affiliation{$^{4}$Max-Planck-Institut f\"ur extraterrestrische Physik, Giessenbachstr. 1, 85748 Garching, Germany}
\affiliation{$^{5}$Instituto de Astrof\'isica de Canarias (IAC); Departamento de
Astrof\'isica, Universidad de La Laguna (ULL), 38200, La Laguna,
Tenerife, Spain}
\affiliation{$^{6}$School of Physics, HH Wills Physics Laboratory, University of Bristol, Tyndall Avenue, Bristol, BS8 1TL, UK}

\begin{abstract}
    The first Euclid Quick Data Release (Q1) provides extensive imaging and spectroscopic data for hundreds of millions of photometric objects across several deep fields. Accurate classifications and photometric redshifts (photo-$z$) for these sources are crucial to maximizing the value of these data. In this work, we perform source classification and photo-$z$ estimation for the Euclid Deep Field North (EDF-N) around the North Ecliptic Pole, using a deep learning framework (DeepDISC) that learns and infers using 9-band images simultaneously. We train three dedicated models for (1) source detection and classification (Model 1), (2) galaxy photo-$z$ (Model 2), and (3) quasar photo-$z$ (Model 3). The Euclid Q1 input source catalog, and classifications and spectroscopic redshifts (spec-$z$) from the Dark Energy Spectroscopic Instrument Data Release 1 are adopted as our training data. DeepDISC source detection achieves overall completeness of ${\sim}93\%$ and purity of ${\sim}80\%$ if using the Euclid source catalog as the ground truth. Using a JWST source catalog within EDF-N as the reference, we estimate a true purity of $\sim 90\%$ for DeepDISC sources. About $99.2\%$, $99.0\%$, and $84.8\%$ of stars, galaxies, and quasars, respectively, are correctly recovered with their spectroscopic classifications. The DeepDISC photo-$z$s show good agreement with spectroscopic redshifts, for both galaxies and quasars. Comparisons with other Euclid Q1 products demonstrate that DeepDISC provides comparable or improved performance in source detection/deblending, classification and photo-$z$, especially for quasars. These results demonstrate the potential of pixel-level deep learning approaches for large-scale sky surveys such as Euclid and Roman, which will continue to improve with better training labels. We release the full DeepDISC source catalog ($\sim 13$~million objects) for EDF-N with classifications and photo-$z$s, including photo-$z$ probability distributions.
\end{abstract}



\maketitle

\section{Introduction}
\label{sec:intro}

Space-based wide-field imaging and spectroscopic surveys such as Euclid \citep{Euclid} are starting to revolutionize the field of astronomy. The first quick data release (Q1) from Euclid \citep{Euclid:Data} provided an extensive set of multi-band (ground-based optical and Euclid) images, grism spectra and catalogs for the Euclid Deep Fields. These early data products demonstrate the capabilities of Euclid and the significant improvements over ground-based wide-area surveys. 

To facilitate the primary science goals of Euclid, the Euclid mission has developed the PHZ processing function to perform source classification, photometric redshift estimation \citep{Salvato2019,Newman2022}, and measurements of galaxy physical properties (\citet{Tucci25}; TM25 hereafter). The PHZ pipeline provides incredibly rich and valuable data products, which form the foundation for a myriad of science applications. However, there are several places where this process can be improved to enhance the scientific value of these data. The PHZ function classifies sources into stars, galaxies, and quasars (QSOs). It delivers a high-purity star sample and a highly complete galaxy sample. However, quasar identification remains challenging, as QSOs are difficult to distinguish from galaxies, and the classification pipeline can be further refined. Additionally, the PHZ photometric redshifts are primarily designed to support Euclid's core science goals, which focus on galaxies. Consequently, PHZ classification and photometric redshifts are not optimal for QSOs.

Pixel-level deep learning algorithm is a powerful method to conduct source detection, deblending, classification, and photo-$z$s \citep{Xu2024,Ting2025,Mendoza2026}. Such approaches have been increasingly adopted in large astronomical surveys and have demonstrated strong performance \citep[e.g.,][]{DIsanto2018,Burke2019,Pasquet2019,Henghes2022,deepdisc:detection,deepdisc:redshift,Roster2024,Roster2025,Siudek2025,Stevens25}. For example, DeepDISC \citep{deepdisc:detection,deepdisc:redshift} is a pixel-level deep learning method, which can provide source detection, deblending \citep[i.e., instance segmentation;][]{Melchior2021}, classification, and photo-$z$ estimation with full posterior probability density functions (PDFs) simultaneously. Using simulated Rubin Observatory Legacy Survey of Space and Time \citep[LSST;][]{LSST:DC2} data, \citet{deepdisc:redshift} validates DeepDISC's performance on source detection, deblending, classification, and photo-$z$ estimation and demonstrates that DeepDISC photo-$z$ can outperform traditional catalog-based (both template-fitting and ML-based) methods for both point estimates and full PDFs. Most recently, \citet{Merz2026} extend DeepDISC to JWST Near Infrared Camera (NIRCam) imaging, demonstrating reliable probabilistic photo-$z$ estimation to $z{\sim}8$ on JADES \citep{Eisenstein2023} data and showing that DeepDISC matches or outperforms EAZY \citep{Brammer2008} template fitting when the same photometric filters are available.

In this work, we train DeepDISC models on real EDF-N images to produce an enhanced catalog with improved source classification and photometric redshifts. This paper is organized as follows. In Section~\ref{sec:method}, we describe the data products used for training and the corresponding training strategies. In Section~\ref{sec:res}, we evaluate the performance of the model predictions and present the final enhanced catalog for EDF-N. In Section~\ref{sec:discuss}, we compare the DeepDISC classification and photometric redshifts with other Euclid results. Finally, we summarize this work in Section~\ref{sec:con}.

\section{Methodology}
\label{sec:method}

\subsection{Data}
\label{sec:data}

The Euclid Deep Field North (EDF-N) is a circular field covering approximately $20\,\mathrm{deg}^2$ around the ecliptic north pole, reaching a depth comparable to that of a single visit of the Euclid Wide Survey in the Q1 data release \citep{Euclid:Data}. The EDF-N Q1 data release provides multi-wavelength images and photometry by combining Euclid the VISible instrument (VIS) and the Near-Infrared Spectrometer and Photometer (NISP) observations with external ground-based data from the Ultraviolet Near Infrared Optical Northern Survey \citep[UNIONS;][]{UNIONS}. Specifically, the Euclid observations include $\Ie$, $\Ye$, $\Je$, and $\He$ bands, while UNIONS contributes Canada-France-Hawaii Telescope (CFHT) $u$, $r$, Hyper Suprime-Cam (HSC) $g$, $z$, and Pan-STARRS $i$ imaging. 

The Euclid MERge Processing Function \citep[MER PF;][]{Euclid:MER} calibrates the multi-wavelength data {in units of predefined regions of the sky, referred to as “tiles.”} It produces a suite of data products, including the MERge catalog (MER) with calibrated photometry for all detected sources, background-subtracted mosaicked images in the nine bands listed above, and the corresponding segmentation maps. Within each tile, the multi-band images are uniformly rescaled to a common pixel scale by the MER PF. In this work, we utilize the 9-band background-subtracted mosaicked images in EDF-N from the Euclid Q1 data release. For GPU processing, {the EDF-N multi-band images in each tile} are divided into $800 \times 800$ pixel patches for training and applications.  

DeepDISC is a supervised deep-learning method and therefore requires ground-truth labels for source detection, classification, and photo-$z$ estimation. For detection training, we adopt the segmentation maps provided by the Euclid MER PF as the ground truth. We apply a simple quality cut, \texttt{spurious\_flag = 0}, to remove spurious detections. We do not impose additional quality cuts in order to preserve detection completeness during training. More stringent selections (e.g., magnitude cuts) can be applied to the model-produced catalog if higher-purity detections are desired. The MER PF segmentation maps are derived from the Euclid $\Ie$ band for optically detected sources; for sources not detected in the optical, the segmentation is constructed from the stacked NIR bands ($\Ye$+$\Je$+$\He$).

For classification and photo-$z$ training, we adopt spectroscopic classifications and redshifts from Dark Energy Spectroscopic Instrument Data Release 1 as ground truth \citep[DESI DR1,][]{DESI:DR1}. We cross-match the Euclid Q1 MER catalog for EDF-N, after applying a quality cut of \texttt{spurious\_flag = 0}, with the DESI DR1 catalog using a matching radius of $0\farcs5$. After cross-matching, we further apply several criteria to select unique, non-sky targets with reliable redshift measurements and no known fitting issues in DESI DR1: 

\begin{itemize}
  \setlength\itemsep{0pt}
  \renewcommand\labelitemi{-}
  \item \texttt{objtype = "TGT"};
  \item \texttt{zcat\_primary = 1};
  \item \texttt{zwarn = 0};
  \item \texttt{coadd\_fiberstatus = 0};
  \item \texttt{deltachi2 > 10}.
\end{itemize}

\begin{figure*}
\centering
\includegraphics[width=\textwidth]{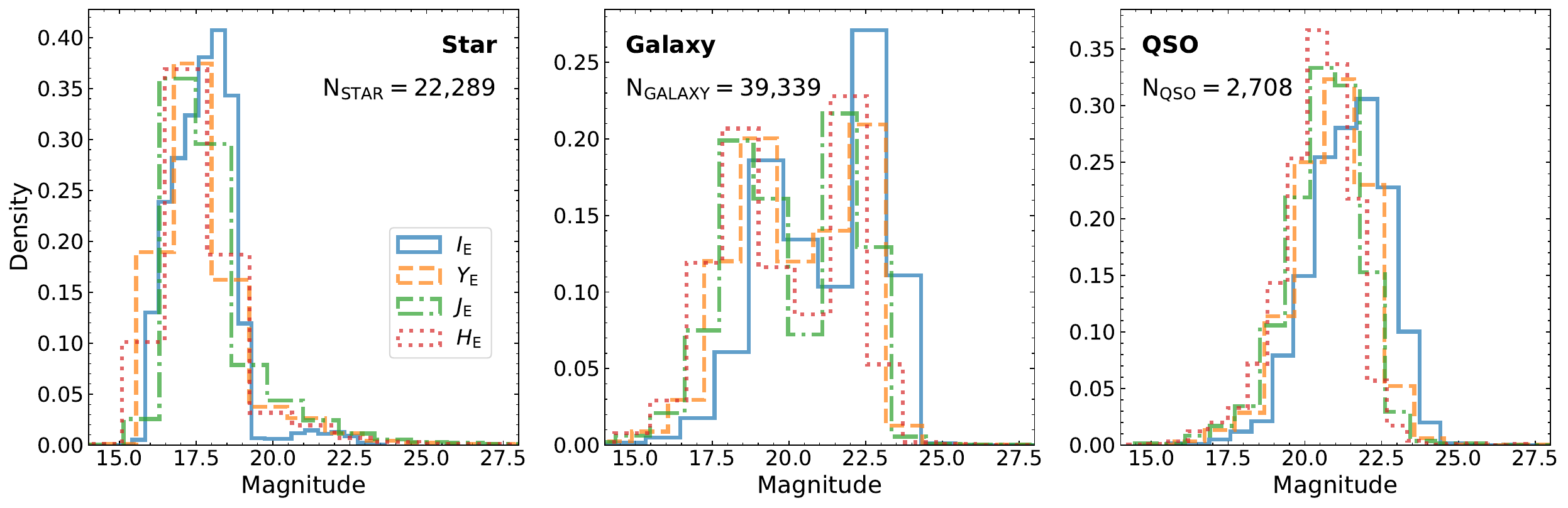}
 \caption{Magnitude (total flux) distributions in Euclid bands ($\Ie$, $\Ye$, $\Je$, and $\He$) for spectroscopically confirmed stars, galaxies, and QSOs, matched between the Euclid MER catalog in EDF-N and DESI DR1. Total fluxes in each band are computed from the 2FWHM aperture flux in the Euclid MER catalog, following the Photometry Cookbook \citep{Euclid:Data}, and converted to magnitudes for display. }
\label{fig:mag_dist}
\end{figure*}

\begin{figure}
\includegraphics[width=0.45\textwidth]{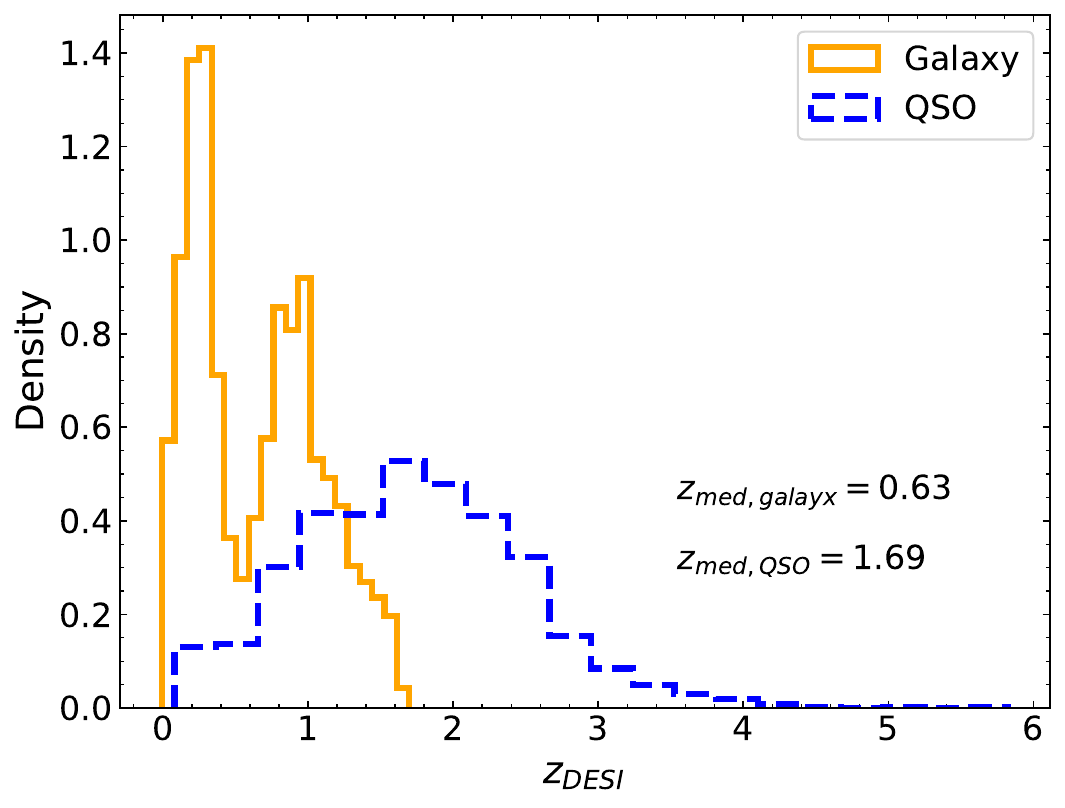}
 \caption{Redshift distributions of galaxies and QSOs matched between the Euclid MER catalog in EDF-N and DESI DR1.}
\label{fig:z_dist}
\end{figure}

The cross-matching results in 64,336 matched sources between EDF-N Q1 and DESI DR1, including 22,289 stars, 39,339 galaxies, and 2,708 QSOs (i.e., unobscured broad-line quasars). The magnitude and redshift distributions for the matched sources are shown in Figures \ref{fig:mag_dist} and \ref{fig:z_dist}. Here, the magnitude in each Euclid band corresponds to the total flux derived by scaling the 2FWHM aperture flux reported in the Euclid MER catalog \citep[see the Photometry Cookbook;][]{Euclid:Data}.

After establishing the ground truth table, we further split the data into three subsets for model training, validation, and testing. The split strategy is determined by the number of sources matched to DESI, in order to optimize the classification and photo-$z$ training, for which the available DESI ground truth is relatively limited, whereas the much larger number of Euclid sources is sufficient for detection training. Specifically, we count the number of DESI-matched sources in each image cutout and assign cutouts to the training, validation, and test sets in descending order of this count. This can improve training efficiency for the image-based DeepDISC model by maximizing the number of DESI-labeled sources per image and reducing unnecessary image loading. Additionally, because sources located near cutout boundaries may appear in multiple cutouts, we define the data split in terms of the number of ``unique" DESI-matched sources, adopting a ratio of 4:1:1 for the training, validation, and test sets. This choice is sufficient for model training and evaluation, as demonstrated in the following sections. The training set contains 9,051 cutouts and 40,002 unique DESI-matched sources. The validation set contains 4,438 cutouts and 10,001 unique DESI-matched sources that do not appear in the training set. The test set consists of 11,697 cutouts and the remaining 14,333 unique DESI-matched sources that are not included in either the training or validation sets. The magnitude and redshift distributions of the three data sets are consistent with those shown in Figures \ref{fig:mag_dist} and \ref{fig:z_dist} for the full sample, indicating that this splitting strategy does not introduce significant biases.

\subsection{Models}
\label{sec:model}

DeepDISC is developed by \citet{deepdisc:detection}, building on the astronomical instance segmentation framework of \citet{Burke2019}, and is implemented in PyTorch \citep{pytorch} leveraging the Detectron2 framework \citep{detectron2}. In this work, we select a backbone called Swin Transformer \citep{swin} with a feature pyramid architecture. The feature pyramid network extracts features at multiple spatial resolutions from the input image, improving sensitivity to objects of different sizes. The feature maps are then passed to the Region Proposal Network (RPN) to generate region proposals about object locations, and the features corresponding to each proposal are then pooled and processed by the Region of Interet (ROI) heads. The ROI heads comprise multiple task-specific networks that take region-level features as input and perform instance-inference on object properties, e.g., segmentation maps, classification, and photo-$z$s. A diagram of the DeepDISC architecture is shown in Figure 3 of \citet{deepdisc:redshift}. 

\begin{figure*}
\centering
\includegraphics[width=0.9\textwidth]{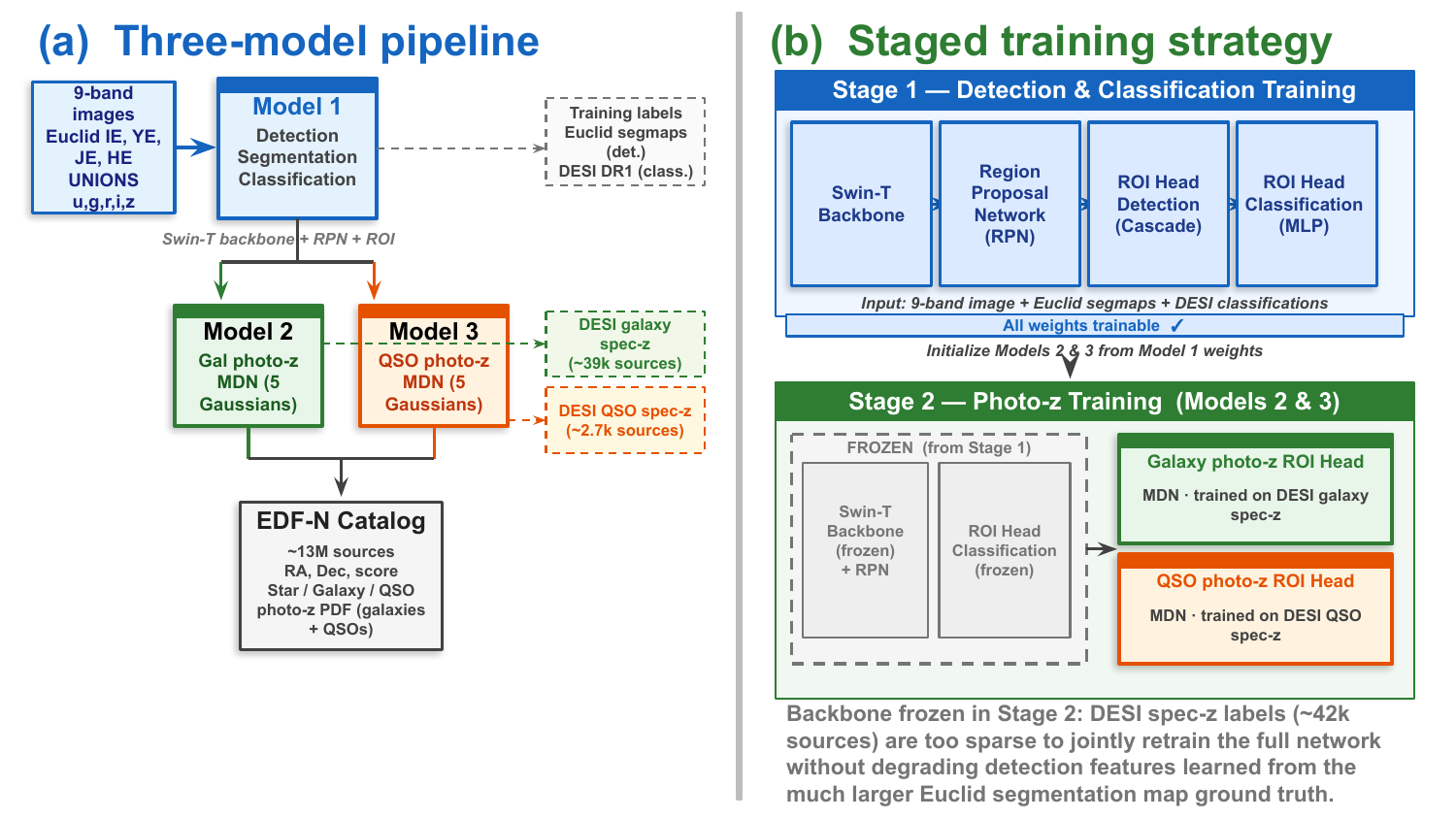}
\caption{(a) The DeepDISC three-model pipeline applied to the Euclid Deep Field North. Nine-band Euclid+UNIONS images are processed by Model 1 for simultaneous detection, instance segmentation, and star/galaxy/QSO classification. Models 2 and 3 append galaxy and QSO photometric redshift PDFs respectively to produce the final merged catalog of $\sim$13 million sources. Dashed lines indicate training label inputs. (b) Staged training strategy. Stage 1 trains all network weights jointly using Euclid segmentation maps and DESI DR1 spectroscopic classifications. Stage 2 initializes Models 2 and 3 from the Stage 1 checkpoint, then freezes the backbone and classification weights and trains only the new MDN photo-$z$ ROI heads on the DESI spec-$z$ sample. Freezing is necessary because the $\sim$42,000 DESI-labeled sources available for photo-$z$ supervision are too sparse to jointly retrain the full network without degrading the detection and deblending features learned from the much larger Euclid segmentation map ground truth.}
\label{fig:workflow}
\end{figure*}

In this work, we train three models with task-specific ROI heads (described below) for source detection and classification, galaxy photo-$z$ estimation, and QSO photo-$z$ estimation, respectively. We found that training these separate models outperforms a single model that trains on all physical labels. Table \ref{tbl:model} summarizes the training strategy adopted in this work. A schematic overview of the three-model pipeline and staged training strategy is shown in \autoref{fig:workflow}. Detailed descriptions of the training design for each model are provided below.

\begin{deluxetable*}{lll}
\tablecaption{Summary of DeepDISC Models and Training Labels\label{tbl:model}}
\tablehead{
\colhead{Model Name} & \colhead{Model Task} & \colhead{Ground Truth}
}
\startdata
Model 1 & Detection \& classification &
\begin{tabular}[t]{@{}l@{}}
Euclid Q1 segmentation map by MER PF \\ in EDF-N
\& DESI classifications
\end{tabular} \\
\hline\\
Model 2 & Galaxy photo-$z$ & DESI DR1 galaxy spec-$z$ \\
\hline\\
Model 3 & QSO photo-$z$ & DESI DR1 QSO spec-$z$ 
\enddata
\end{deluxetable*}

The first model (Model 1 hereafter) focuses on source detection and classification, which is the first step for photo-$z$ measurements. For detection, we use the cascade ROI head \citep{cascade}, which works well in previous DeepDISC work \citep{deepdisc:detection, deepdisc:redshift}. A cascade ROI head refines detections through multiple detection stages by progressively enforcing stricter agreement with the ground-truth annotations, i.e., Euclid segmentation maps in this work, leading to higher-quality detection results. For classification, we use a basic multilayer perception (MLP) with cross-entropy loss function. We convert the classification annotations to probability vectors, where each element corresponds to the target probability of a given class. We take the DESI classification as the ground truth; therefore, the probability of the corresponding class is set to 1, while those of the other classes are set to 0. Only sources with available DESI classification labels are included in the training and loss calculation. 


Model 1 is trained with two ROI heads mentioned above together and can therefore infer the instance segmentation maps and the classification probability vectors for a given image. The model is initialized from the checkpoint provided by \cite{li2022exploring}, which was pretrained on the ImageNet21k and MS-COCO datasets \citep{deng2009imagenet,MSCOCO} of everyday RGB images. The top layer of the model is not copied from the checkpoint due to the difference in channel numbers for RGB vs Euclid images, and a few layers in the ROI heads are not copied due to differences in the number of object classes. Despite this discrepancy, the remaining copied layers still provide a helpful starting point for the model, as demonstrated in \cite{deepdisc:detection} and \cite{deepdisc:redshift}. We start with a learning rate of 0.0001 and train the entire network without freezing of the backbone weights. We train the model for 30 total epochs and decrease the learning rate by a factor of 10 each time at 10 and 20 epochs. These parameters are empirically selected based on the behavior of the loss function during training. During training, we apply standard spatial augmentation including random horizontal and vertical flips. No contrast or color augmentation is applied, as the pixel scaling of the 9-band Euclid+UNIONS images is fixed by the MER pipeline. The learning rate is reduced because we observe sudden increases in the loss around the 10th and 20th epochs. Such spikes are typically caused by an excessively large learning rate, which can lead to unstable updates and overshooting of the local minimum during optimization. After adjusting the learning rate, the model was also trained for an additional 30 epochs, during which the loss remained stable.


The second and third models (hereafter Models 2 and 3) are specifically designed to estimate photo-$z$s for galaxies and QSOs, respectively. We incorporate a Mixture Density Network (MDN; \citealt{MDN}) into the ROI heads and train the models accordingly. A detailed description of this network architecture and its feasibility, based on simulated LSST data \citep{LSST:DC2}, is presented in \citet{deepdisc:redshift}. 
Briefly, the MDN outputs a redshift PDF for each object, modeled as a mixture of five Gaussian components, following \citet{DIsanto2018} and \citet{deepdisc:redshift}, who show this parameterization performs well for galaxy and quasar images.



Models 2 and 3 are initialized from Model 1, which is trained for source detection and classification. During photometric-redshift training, the backbone and classification weights from Model 1 are frozen. While in principle joint training of detection, deblending, classification, and photo-$z$ is mutually beneficial (e.g., deblending in particular reduces contamination from neighboring sources in the photo-$z$ estimate), simultaneous fine-tuning of all weights is limited here by the relatively small number of DESI-labeled sources available for photo-$z$ supervision. Jointly updating the backbone on this limited labeled sample risks degrading the well-trained detection and deblending features learned from the much larger Euclid segmentation map ground truth. We therefore adopt a staged training strategy as a practical compromise for Euclid Q1 (\autoref{fig:workflow}b). As spectroscopic coverage grows with Euclid DR1 and programs such as C3R2 \citep{Masters2017C3R2,Stanford2021} and 4C3R2 \citep{Guglielmo2020,Gruen2023}, a fully joint training regime will become feasible and is planned for future work. As illustrated in \autoref{fig:workflow}b, the current staged strategy isolates the photo-$z$ ROI heads from the backbone during Stage 2 training precisely to protect the detection and deblending features learned in Stage 1.
The new ROI heads are trained for an additional 30 epochs. The learning rate is also initialized at 0.0001 and reduced by a factor of 10 at epochs 10 and 20. Model 2 is trained exclusively on DESI galaxy redshifts, such that only sources classified as galaxies by DESI are included in the training, while all other sources are masked. On the other hand, Model 3 is trained exclusively on DESI QSO redshifts. Combined with the classifications provided by Model 1, Models 2 and 3 are expected to provide adequate photo-$z$ estimates for galaxies and QSOs, respectively, from a given image.

The models are trained using 8 NVIDIA H200 GPUs, with total training times of approximately 5.5, 3.5, and 3.0 hr for Models 1–3, respectively. We then apply the models to the test sets to evaluate their performance, which takes about 1.4 hr per model on a single NVIDIA A100 GPU. The models are also applied to the full EDF-N images, with total inference and I/O times of 11.30, 8.78, and 8.46 hours for Models 1–3, respectively, on a single NVIDIA A100 GPU, corresponding to an inference rate of $\sim 10^6$ objects per hour.

\section{Results}
\label{sec:res}

After training, we apply the models to the test set for inference and evaluate their performance. No test set images were seen during training. Below we demonstrate the performance of DeepDISC on source detection, classification and photo-$z$ estimation, and present the full DeepDISC catalog in Section~\ref{sec:cat}. We defer the caveats of our deep learning approach to Section~\ref{sec:caveats}. Because the current DeepDISC implementation does not output source centroid and photometry directly, we follow simple but reasonable recipes (detailed below) to calculate source position and fluxes, based on the output segmentation map for each source.

\subsection{Detection}
\label{sec:det}

\begin{figure}
\centering
\includegraphics[width=0.5\textwidth]{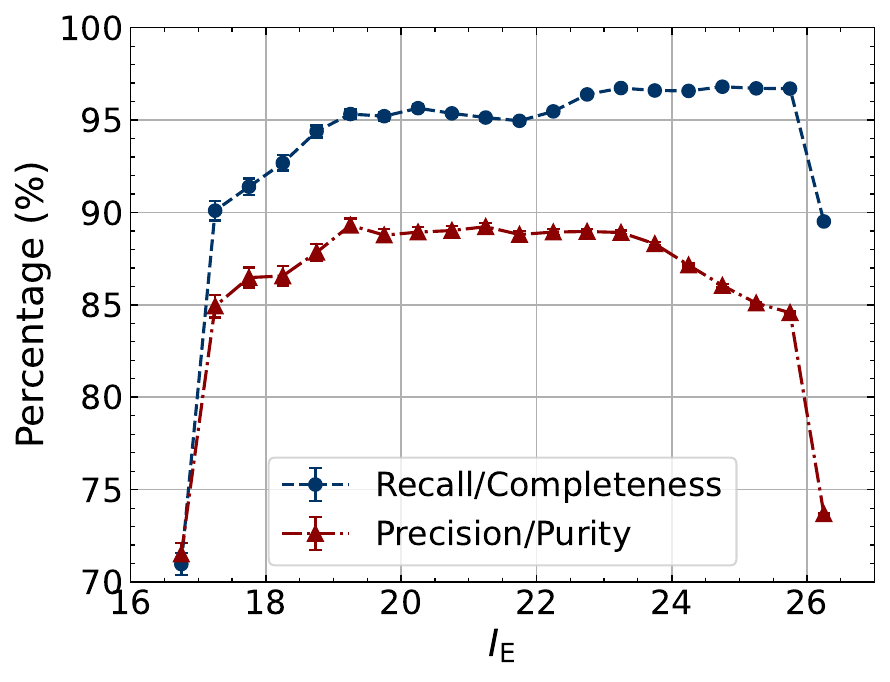}
 \caption{Detection recall (completeness) and precision (purity) of Model 1 as functions of $\Ie$ on the test set. Blue circles denote recall, and red triangles denote precision. Poisson uncertainties are shown.}
\label{fig:det_rp}
\end{figure}

\begin{figure*}
\centering
\includegraphics[width=\textwidth]{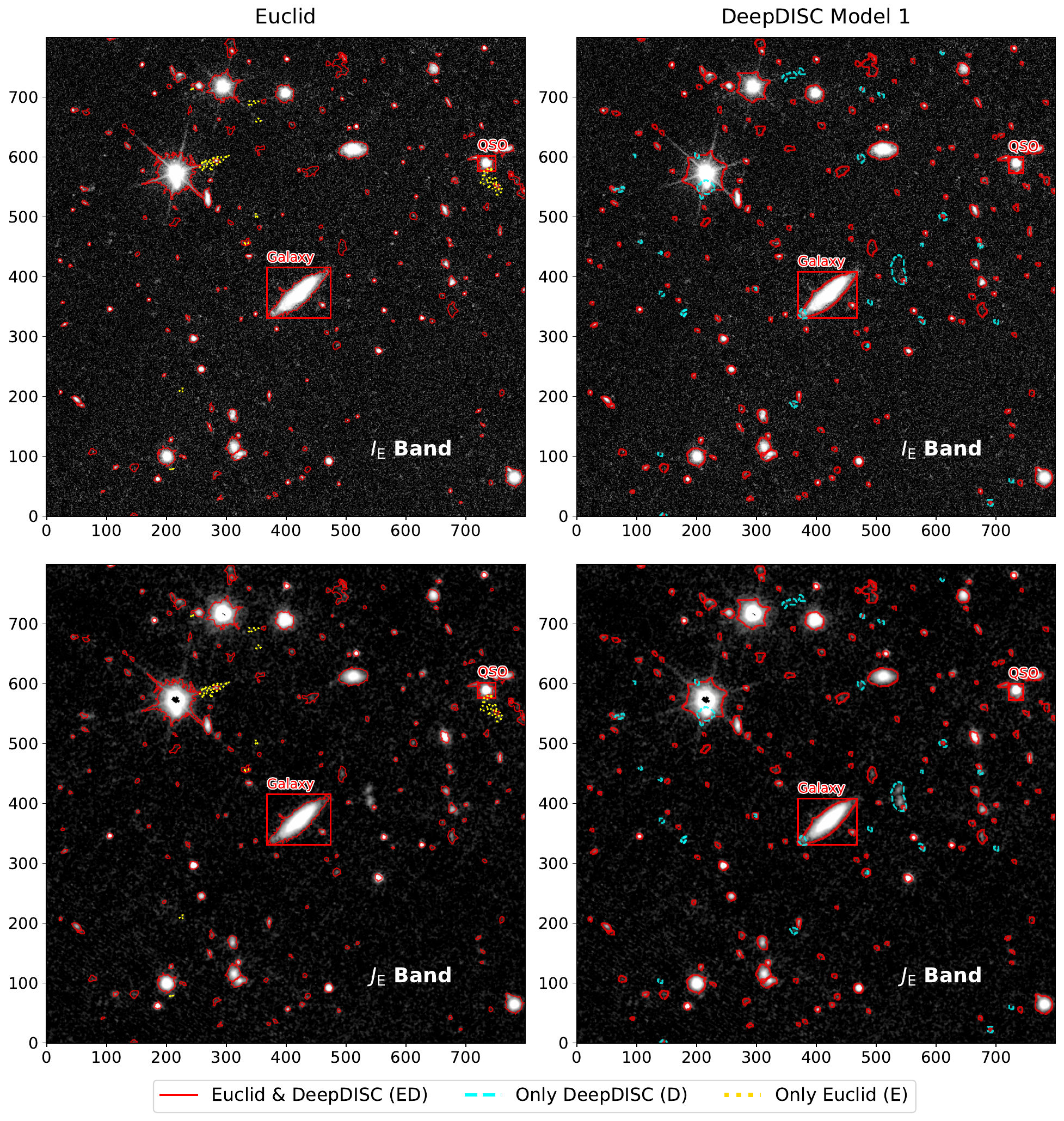}
\caption{Example comparisons between Euclid (left) and Model 1 (right) segmentation maps. The Euclid segmentation maps are shown in the left panels, while the Model 1–predicted segmentation maps are shown in the right panels. The top row displays cutouts in the $\Ie$ band, and the bottom row shows cutouts in the $\Ye$ band. Red solid contours indicate objects detected by both Euclid and Model 1 (ED); blue dashed contours indicate objects detected only by the DeepDISC Model 1 (D); and yellow dotted contours indicate objects detected only by Euclid (E). DESI objects are additionally highlighted with red squares, with classifications from both DESI and Model 1 indicated. Overall, most model detections are consistent with Euclid detections. Unmatched detections primarily arise from faint sources, artifacts, blended systems, or sources that are not visible in all bands. }
\label{fig:seg}
\end{figure*}

We define the recall (i.e., completeness) and precision (i.e., purity) to quantitatively evaluate the performance of detection by Model 1 against the benchmark (in this case, the Euclid MER catalog):

\begin{equation}\label{eq:r}
    r = \rm \frac{ED}{ED+E},
\end{equation}

\begin{equation}\label{eq:p}
    p = \rm \frac{ED}{ED+D},
\end{equation}
where we use three detection counts ED, E, and D, corresponding to sources detected by both Euclid and DeepDISC, those only detected by Euclid, and those only detected by DeepDISC, respectively. Here we assume all sources in the MER catalog are real. In practice, however, some E objects are bogus sources and some D objects are genuine sources missed by the MER catalog, and vice versa (see below). Cross-matching between the Euclid MER catalog and the Model 1 catalog is based on the center of the bounding box\footnote{Using the bounding box center for the cross-match is the simplest approach and works for the vast majority of sources. More sophisticated definitions of source centroid can be used. However, such definitions have ambiguities for sources with complex morphologies and can be band dependent. In our final catalog products, we include information for source segmentation maps and the user can calculate the desired source centroid in any given band.} with a matching radius of 1.5 arcsec. Only the nearest counterparts are selected. This matching scheme works for the vast majority of sources except for the brightest and/or most extended objects, which is further discussed below.


Model 1 detection performs well on the test set, with overall recall of $93\%$ and precision of $80\%$ for the full sample (dominated by objects towards the faint end). The fractions are higher if we exclude the brightest and faintest objects. The recall and precision as a function of $\Ie$ magnitudes are shown in Figure \ref{fig:det_rp}. For the purpose of comparisons, the magnitudes used here are computed from the total flux within the Euclid MER segmentation map for Euclid-detected objects (ED or E), and from the DeepDISC segmentation map for objects detected only by Model 1 (D); this definition is adopted throughout the following comparisons. The recall is higher than $95\%$ at $19<\Ie<26$, and the precision is higher than 85\% at $17.5<\Ie<25.5$. 

Both recall and precision drop at the bright and faint ends. To investigate these behaviors we perform visual inspection on discrepant detection cases (i.e., E and D), which directly impact the recall and precision following Equations \ref{eq:r} and \ref{eq:p}. Bear in mind that a detection labeled as E when cross-matching the model with the MER catalog does not necessarily correspond to a missed real astronomical object; similarly, a detection labeled as D may still correspond to a genuine source. This is because the detection ground truth is derived from the Euclid segmentation map, which may itself include spurious detections or missed real objects. We therefore visually inspected 100 randomly selected image cutouts. In these 100 cutouts, there are 20,556 EDs, 1,624 Es, and 3,407 Ds. We show some examples in Appendix~\ref{appx1} and summarize the main findings below.


\begin{itemize}
    \item At the bright end, discrepancies between Euclid and DeepDISC detections are primarily driven by offsets in the predicted source centroids for bright stars and very extended nearby galaxies (Figure~\ref{fig:ext}). Visual inspection shows that all Es and Ds at $\Ie < 19$ are associated with such objects. DeepDISC adopts the bounding-box centroid as the source position. However, the center of the input bounding box derived from the Euclid segmentation map is sometimes biased by the artifacts (e.g., spikes). Recovering accurate bounding boxes in such cases is challenging and can result in mismatches (left column of Figure \ref{fig:ext}). Additionally, blending around bright sources (middle and right panels) further contributes to inconsistent detections between Euclid and DeepDISC. Overall, these objects are still included in the DeepDISC catalog in EDF-N, albeit their fiducial positions may be inaccurate.
    
    \item At the faint end, the decreasing recall and precision are mainly due to unreliable detections with low signal-to-noise ratios (SNR), given that the average $5\sigma$ depth in the $\Ie$ band is about 26 \citep{Tucci25}. Except for random low-SNR detections, faint artifacts, such as suspicious detections around stellar diffraction spikes (e.g., left panel of Figure \ref{fig:artifact}), imperfect background subtraction, or stacking artifacts (e.g., right panel of Figure \ref{fig:artifact}), contribute to about 20\% (231/1283) of Es and 8.3\% (253/3017) of Ds at $\Ie > 24$. Some of these artifacts can be relatively bright (e.g., middle panel of Figure \ref{fig:artifact}), but a magnitude cut can remove most of artifacts as well as random low-SNR detections.

    \item Blended sources in either Euclid or DeepDISC detections can lead to discrepancies between the two catalogs. {Visual inspection indicates that multiple objects are sometimes grouped within a single segmentation map produced by the Euclid MER catalog \citep[see deblending strategy in][]{Euclid:MER} or DeepDISC, biasing the bounding box centers. In such cases, successful separation of these objects by the other method can lead to mismatches.} As we do not refine the Euclid MER catalog deblending, the model is not trained on fully deblended segmentation maps, which may lead to occasional good or bad deblending by DeepDISC. DeepDISC appears to perform better at separating objects near stellar spikes (e.g., left panel of Figure \ref{fig:deblend}). However, in most situations, successful deblending appears to occur somewhat randomly in both Euclid and DeepDISC (e.g., middle and right panels of Figure \ref{fig:deblend}). {We note, however, that in cases where the blended sources have different colors, DeepDISC is more likely to deblend them because it uses multi-band images simultaneously in detection (Figure \ref{fig:multi_deblend})}. 
    
    \item Multiple DeepDISC detections at nearly the same position can also lead to DeepDISC-only detections. Most of such D cases are very faint in the $\Ie$ band, with magnitudes fainter than 24 (Figure \ref{fig:multi_spurious}). About 9.2\% (279/3017) of Ds at $\Ie > 24$ show this behavior. Such cases can arise when DeepDISC produces spurious detections due to noise fluctuations. As shown in the upper panels of Figure~\ref{fig:multi_spurious}, these detections are typically faint in all bands and are likely bogus detection by DeepDISC. However, some deblended objects exhibit markedly different morphologies across bands, leading to two closely separated detections (e.g., lower panels of Figure \ref{fig:multi_spurious}), where one component is brighter in certain bands. We therefore conclude that multiple detections by DeepDISC at nearly the same position that are faint in all bands are likely spurious, whereas close pairs with distinct colors may correspond to real sources or band-dependent morphologies identified by DeepDISC but not by Euclid.
\end{itemize}

\subsection{Classification}
\label{sec:class}

\begin{figure*}
\centering
\includegraphics[width=\textwidth]{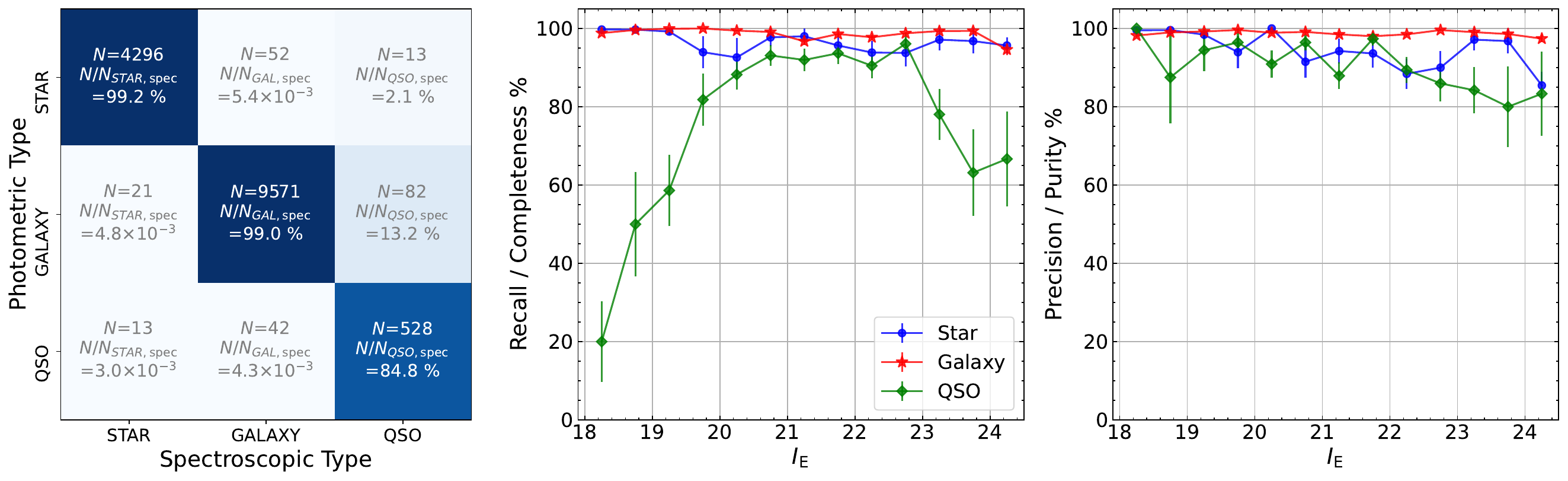}
 \caption{DeepDISC classification results on the test set. The left panel shows a confusion matrix comparing DeepDISC predictions with DESI spectroscopic classifications. Each cell reports both the number and fraction of DESI objects assigned to each category by DeepDISC. Color intensity indicates the fraction in each cell, and the dark diagonal signals demonstrate that most DeepDISC classifications are consistent with DESI labels. The middle and right panels present the recall (completeness) and precision for each class as a function of \Ie, respectively. Definitions of the DeepDISC classification recall and precision are provided in Section~\ref{sec:class}. Blue circles, red stars, and green diamonds denote stars, galaxies, and QSOs, respectively. }
\label{fig:class_dd}
\end{figure*}

For classification evaluation, only objects detected by both Euclid and Model 1 (EDs) with DESI spectroscopic classifications are included. As described in Section \ref{sec:model}, Model 1 predicts class probabilities for stars, galaxies, and quasars. We assign the class with the highest probability as the final classification, which is a strict criterion but delivers good performance. Statistics of the classification performance are shown in Figure~\ref{fig:class_dd} in terms of the recall and precision metrics. The definitions of classification recall and precision are analogous to those for detection, but use DESI DR1 classifications as the ground truth.

The left panel of Figure~\ref{fig:class_dd} presents the confusion matrix of the Model 1 classification results, which show good agreement for all three types (star, galaxy, and quasar). Firstly, the star and galaxy classifications show excellent agreement with DESI results. More than 99\% of DESI stars and galaxies are correctly classified as stars and galaxies, respectively. On the other hand, the overall precision for Model 1 stars and galaxies reaches 98.5\% and 98.9\%, respectively, indicating that objects predicted by the model as stars or galaxies are mostly likely to be true stars or galaxies\footnote{This purity is measured against the composition (stars, galaxies and QSOs) of the full DESI spectroscopic sample. If a real photometric sample is overwhelmingly dominated by stars, even a tiny fraction of misclassifications could lead to significant contamination in samples of other classes.}. Secondly, although the quasar classification is not as good as that for stars and galaxies, Model 1 achieves reasonably high completeness and precision: 84.8\% DESI quasars are properly classified as quasars by Model 1, while 90.6\% of the model classified quasars are indeed quasars. 


The completeness and precision are further evaluated in different $\Ie$ magnitude bins, as shown in the middle and right panels of Figure~\ref{fig:class_dd}. Model 1 achieves high completeness and precision ($\gtrsim 90\%$) for stars and galaxies across all $\Ie$ bins. The quasar classification precision also remains high over the full $\Ie$ range, with a modest decline at $\Ie \gtrsim 22$, indicating that objects identified as quasars by the model are generally reliable. Moreover, the quasar classification completeness also remains high ($>80\%$) over the range $19.5 < \Ie < 22.5$, but drops significantly at both the bright and faint ends. The reduced completeness and precision toward the faint end is primarily attributed to lower SNRs for intrinsically faint objects, or to incomplete structures near cutout boundaries. In contrast, the reduced completeness at the bright end may result from confusion between bright host galaxies and their central AGNs, e.g., some low-redshift quasars are classified by Model 1 as galaxies due to the significant host galaxy emission and extended morphology. Visual inspection of objects classified as quasars by DESI only (but not by DeepDISC Model 1) in the test set reveals that many of these sources are indeed at low redshift and exhibit strong host galaxy contributions, leading to their misclassification as galaxies. Of the 95 such objects for which DESI provides redshifts, 40\% lie at $z <0.5$, where host galaxy contamination is most severe. This limitation is partly a consequence of the relatively small number of low-$z$ broad-line AGN in the DESI training sample (\autoref{fig:z_dist}), and will improve as larger samples of low-$z$ training set become available in Euclid DR1.



\subsection{Photometric Redshifts}
\label{sec:z}

\begin{figure*}
\centering
\includegraphics[width=\textwidth]{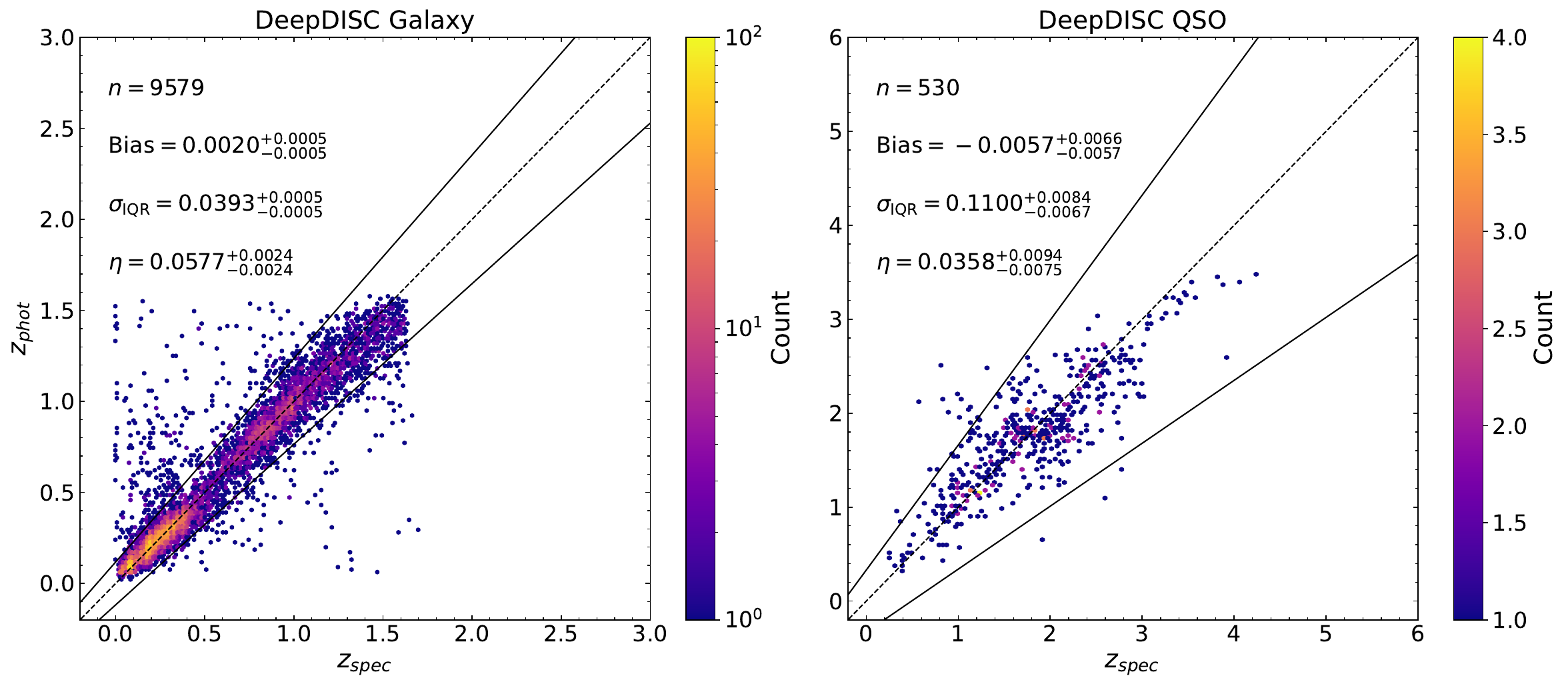}
 \caption{The 2D histograms of DeepDISC photo-$z$, $z_{\rm phot}$, versus DESI spec-$z$, $z_{\rm spec}$, for objects identified as galaxies or quasars by both DESI and the models in the test set. The left panel shows Model 2 predicted $z_{\rm phot}$, trained only with DESI galaxy $z_{\rm spec}$, which shows good agreement with DESI results. The right panel displays the comparison between DESI quasar $z_{\rm spec}$ and $z_{\rm phot}$ predicted by Model 3, trained only with DESI quasar $z_{\rm spec}$. The color bar indicates the number of objects per hexagonal bin. The color scale is logarithmic (linear) for the galaxy (QSO) panel. The DeepDISC quasar $z_{\rm phot}$ also shows a good linear relation with $z_{\rm spec}$, but with larger scatter compared to the galaxy case, due to the smaller number of quasars with ground-truth redshifts (DESI $z_{\rm spec}$). The total number of objects, median bias relative to $z_{\rm spec}$, scatter $\sigma_{\rm IQR}$, and outlier fraction $\eta$, defined in Section~\ref{sec:z}, are reported in each panel.}
\label{fig:z_dd}
\end{figure*}

Although a single model can be trained to detect and classify sources, as well as providing photo-$z$ estimates, we found that training separate models of photo-$z$ greatly improves the accuracy of photometric redshifts. Model 2 and Model 3 are trained to estimate photo-$z$s for galaxies and quasars, respectively. Both models are applied to the test set for inference, and their photo-$z$ predictions are compared with DESI spec-$z$s. To avoid confusion arising from mismatched classifications between the models and DESI, the following statistics are restricted to objects with consistent classifications in both DeepDISC and DESI catalogs. 


Model 2 and Model 3 provide the full PDF of photo-$z$ ($z_{\rm phot}$). We adopt the median of the PDF as the fiducial $z_{\rm phot}$ estimate, because using median photo-$z$ values usually yields a slightly lower outlier fraction than mode estimates, although these point estimates are generally consistent with each other \citep{Tucci25}. We compute the same statistics following \citet{deepdisc:redshift}. For completeness, we briefly summarize the statistics used for the photo-$z$ comparison below. We quantify the bias $e_z$ to the DESI spec-$z$ ($z_{\rm spec}$) as

\begin{equation}\label{eq:ez}
    e_z = (z_{\rm phot} - z_{\rm spec})/(1+z_{\rm spec}). 
\end{equation}

The scatter, $\sigma_{\rm IQR}$ of the estimates is defined as 

\begin{equation}\label{eq:sigma}
    \sigma_{\rm IQR} = (e_{z75} - e_{z25})/1.349,
\end{equation}
where $e_{z75}$ and $e_{z25}$ are the 75th and 25th percentiles of $e_z$, and the scaling factor of 1.349 is introduced so that the interval spanned by $\sigma_{\rm IQR}$ encloses the same probability as one standard deviation for a standard normal distribution. We also define the outlier fraction as

\begin{equation}\label{eq:eta}
    \eta = N_{\rm out} / N_{\rm tot},
\end{equation}
where $N_{\rm out}$ is the number of galaxies/quasars with $|e_z| > \max (3\sigma_{\rm IQR}, 0.06)$, following the definition of \citet{Schmidt2020}. We note that a small outlier fraction does not necessarily imply good performance if the scatter of photo-$z$ around spec-$z$ is large. The uncertainties of the above statistics are estimated using bootstrap resampling.

A 2D histogram of the comparison between model predicted photo-$z$s and DESI spec-$z$s is shown in Figure \ref{fig:z_dd}. The left panel compares the galaxy photo-$z$s predicted by Model 2 with the DESI spec-$z$s. Model 2 predicted photo-$z$s for 9,579 DESI galaxies in the test set show very good agreement with DESI spec-$z$s, with a median bias $e_{z,{\rm med}}$ of $2.0^{+0.5}_{-0.5}\times10^{-3}$ and a scatter $\sigma_{\rm IQR} = 3.93^{+0.05}_{-0.05}\times10^{-2}$. The outlier fraction, $\eta = 5.77^{+0.24}_{-0.24}\%$, is also very low. The right panel compares the quasar photo-$z$s by Model 3 with the DESI spec-$z$s. Model 3 predicted photo-$z$s for 530 DESI quasars generally agree with spec-$z$s, but with larger absolute value of median bias and scatter compared with those for galaxies. The median bias $e_{z,{\rm med}}$, scatter $\sigma_{\rm IQR}$, and outlier fraction $\eta$ of Model 3 quasar photo-$z$s to DESI spec-$z$s is $-5.7_{-5.7}^{+6.6}\times10^{-3}$, $1.10_{-0.07}^{+0.08}\times10^{-1}$, and $3.58^{+0.94}_{-0.75}\%$, respectively. Considering the difference in the numbers of DESI galaxies and quasars (39,339 versus 2,708) in EDF-N, worse performance for quasar photo-$z$ prediction is as expected. However, we note that current performance for Model 3 is already comparable to or even better than other works (see Section \ref{sec:z_comp}).

\subsection{The DeepDISC catalog}\label{sec:cat}

We present the full catalog in EDF-N from DeepDISC with classifications and photo-$z$ information whenever applicable. The inference over the full EDF-N field is performed on cutouts generated within each tile, after which the catalogs from all tiles are combined. For each tile, the sky area is partitioned into $800\times800$ pixel cutouts, with a 100-pixel overlap applied along the positive-$x$ and positive-$y$ directions. The overlapping regions are designed to mitigate edge effects, particularly unreliable inferences resulting from incomplete segmentation of objects at cutout boundaries. After cutout generation, we apply Models 1–3 to all cutouts, producing three independent catalogs for EDF-N. We adopt the catalog predicted by Model 1 as the baseline and update its redshift information using the catalogs generated by Models 2 and 3. Specifically, galaxies and quasars identified in the Model 1 catalog are cross-matched with the Model 2 and Model 3 catalogs using a matching radius of $0\farcs5$. The corresponding redshift measurements from Models 2 and 3 are then incorporated into the final merged catalog. Objects classified as stars by Model 1, as well as a small fraction of unmatched sources due to the updated weights of RPN for Models 2 and 3, are not assigned redshift information. Descriptions of the catalog columns are provided in Table \ref{tbl:catalog}. The final catalog is available at \url{https://ariel.astro.illinois.edu/euclid/edfn/q1/deepdisc/}, along with the segmentation map edge coordinates for each target, where the coordinates are remapped to the coordinate system of each Euclid tile.


The final DeepDISC catalog contains 13,355,031 distinctive sources. We cross-match the final DeepDISC catalog with the Euclid MER catalog in EDF-N, applying a quality cut of \texttt{spurious\_flag = 0} in MER, consistent with the ground-truth catalog described in Section~\ref{sec:data}. A matching radius of $0\farcs5$ is adopted. The Euclid MER catalog in EDF-N with \texttt{spurious\_flag = 0} contains 10,430,497 sources. We identify 9,713,747 common sources in both catalogs, meaning 93.1\% of the Euclid MER sources are recovered by DeepDISC. 

By number counts, the DeepDISC EDF-N catalog contains $\sim 30\%$ more sources than Euclid MER. This increase in source detection is at least partly due to DeepDISC's improved ability to deblend close companions. To evaluate the true purity of DeepDISC detected sources, we use the NEXUS-JWST catalog \citep{nexus,Zhuang2024} that covers a small ($\sim 400\,{\rm arcmin^2}$) area within EDF-N. The NEXUS JWST imaging reaches sufficient depths to detect all sources up to $\sim 28$~mag, making it an ideal reference sample to estimate the true purity of DeepDISC source detection. Within the NEXUS footprint, 54413/60146 of the DeepDISC sources are matched to JWST sources within 0\farcs5, implying a global purity of about 90\%. Unfortunately the same JWST sample cannot be used to evaluate the true completeness of the DeepDISC catalog because of the very different detection characteristics of the JWST observations. 


\begin{deluxetable*}{ccc}
\tablecaption{The DeepDISC Catalog for EDF-N\label{tbl:catalog}}
\tablehead{\colhead{Number} & \colhead{Column} & \colhead{Description}}
\startdata
1  & pred\_id   & Unique ID of Model~1 detected object \\
2  & tile\_id   & Unique ID of Euclid tile             \\
3  & ra\_bbox        & Right ascension of the center of the bounding box predicted by Model 1 (deg) \\
4  & dec\_bbox     & Declination of the center of the bounding box predicted by Model 1 (deg) \\
5  & type\_prob & Classification probability vector predicted by Model 1 \\
6 & type\_pred\tablenotemark{a} & Object type with maximum probability by Model 1\\
7 & gmm\tablenotemark{b} & Parameters of the five-component Gaussian Mixture photo-$z$ PDF\\
8-10 & z\_med, z\_p16, z\_p84 
       & Median, 16th, and 84th percentiles of the photo-$z$ PDF \\
11-13 & z\_mode, z\_hpd\_lo, z\_hpd\_hi 
       & Mode and 68\% highest posterior density (HPD) credible interval of the photo-$z$ PDF\\
14-22 & mag\_*\tablenotemark{c} & AB magnitudes derived from summed flux within DeepDISC segmentation maps \\
 & & 4 Euclid ($\Ie$, $\Ye$, $\Je$, $\He$) + 5 UNIONS ($ugriz$) bands \\
23-26 & nsave, ng, ni, nj\tablenotemark{d} & Indices for segmentation map retrieval
\enddata
\tablecomments{Detailed usage of this catalog is provided in a python notebook tutorial along with the catalog. 
\tablenotetext{a}{The ``type\_pred'' values of 0, 1, and 2 correspond to stars, galaxies, and quasars, respectively.}
\tablenotetext{b}{The parameters in ``gmm'' are stored as $(\tilde{w}_i, \mu_i, \log \sigma_i)$ for five Gaussian components, where $\exp(\tilde{w}_i)$, $\mu_i$ and $\sigma_i$ are the Gaussian mixture weight, mean, and dispersion for each Gaussian.}
\tablenotetext{c}{These magnitudes are provided for reference only and can be used to remove faint and spurious objects. For more accurate photometry, a dedicated pipeline product should be used.}
\tablenotetext{d}{To store the segmentation maps efficiently, the edge coordinates (remapped onto the Euclid tile coordinate system) are concatenated into 1D arrays, and stored in several separate binary files to save space. Columns 23-26 list the indices to retrieve the segmentation map edge coordinates from these binary files for a given source, which the user can use upon the official Euclid image tiles. More details are included in the provided python tutorial. }}
\end{deluxetable*}

\section{Discussion}
\label{sec:discuss}

\subsection{Comparison with Other EDF-N Catalogs}
\label{sec:comp}

\subsubsection{Classification Comparison}
\label{sec:class_comp}

We first compare the DeepDISC classification results with those from other Euclid classification studies. Source catalogs from \citet{Tucci25} (\citetalias{Tucci25}) and \citet{Stevens25} (\citetalias{Stevens25}) are included in the comparison, as these two catalogs represent the most complete Euclid Q1 catalogs with classifications. To ensure a rigorous comparison, we require the objects have classifications from all of the DESI DR1, DeepDISC and \citetalias{Tucci25}/\citetalias{Stevens25} catalogs. Table \ref{tbl:class_comp} summarizes the number of sources including in each catalog used for classification comparison. 

\begin{table*}[!htbp]
\centering
\caption{Classification Comparison.}
\label{tbl:class_comp}
\begin{tabular}{lcccc}
\hline
\hline
Reference & $N_{\rm matched}$ & $N_{\rm Star,DESI}$ & $N_{\rm Galaxy,DESI}$ & $N_{\rm QSO,DESI}$ \\
\hline
This work (Model 1) & 14,618  & 4,330 & 9,665 & 623 \\
Euclid PHZ (MT25; single flag)       & 12,920 & 3,320 & 9,059 & 559\\
Diffusion Model (GS25)               & 4,835 & 167 & 4,461 & 207 \\
\hline
\end{tabular}
\tablecomments{
Number of objects (in the test set) used to evaluate classification performance. The objects are detected by both Euclid MER catalog and DeepDISC Model 1 in the test set. For the first row, $N_{\rm matched}$ denotes the number of objects matched between DeepDISC and DESI. For the second or third row, $N_{\rm matched}$ represents the number of objects jointly matched among DeepDISC, the corresponding method, and DESI. $N_{\rm Star,DESI}$, $N_{\rm Galaxy,DESI}$, $N_{\rm QSO,DESI}$ present the numbers of DESI-identified stars, galaxies, and quasars within each matched sample.
}
\end{table*}


\citetalias{Tucci25} provides a catalog with classifications, photo-$z$s, and physical properties of galaxies generated by Euclid PHZ processing function, and we refer to it at the Euclid PHZ catalog hereafter. A supervised machine learning method called Probabilistic Random Forest (PRF) is used for classification in the Euclid PHZ catalog. The method is not based on images like DeepDISC, but can also provide the probabilities of star, galaxy, and quasar for an object given the fluxes in different bands. In the PRF classifiers, probability thresholds are applied to improve sample purity and can result in multiple classifications for a single object. In contrast, DeepDISC assigns each object to the class with the maximum probability. Therefore, we only select objects with a single classification in the PHZ catalog, indicating that one class has a dominant probability. 

\citetalias{Stevens25} trained a diffusion model based on Euclid \Ie-band images, so we will refer it as the Diffusion Model catalog hereafter. The model reconstructs the light distribution of normal galaxies by masking the central light distribution of each object. After subtracting the reconstructed profile from the original source image, the reconstruction errors in the central pixels provide a diagnostic to distinguish normal galaxies (i.e., galaxies without an AGN) from other source categories. \citetalias{Stevens25} use \texttt{phz\_star\_prob\_phz\_class $<0.3$} in the Euclid PHZ catalog (\citetalias{Tucci25}) to select quasars from all sources not classified as galaxies. 

\begin{figure*}
\centering
\includegraphics[width=\textwidth]{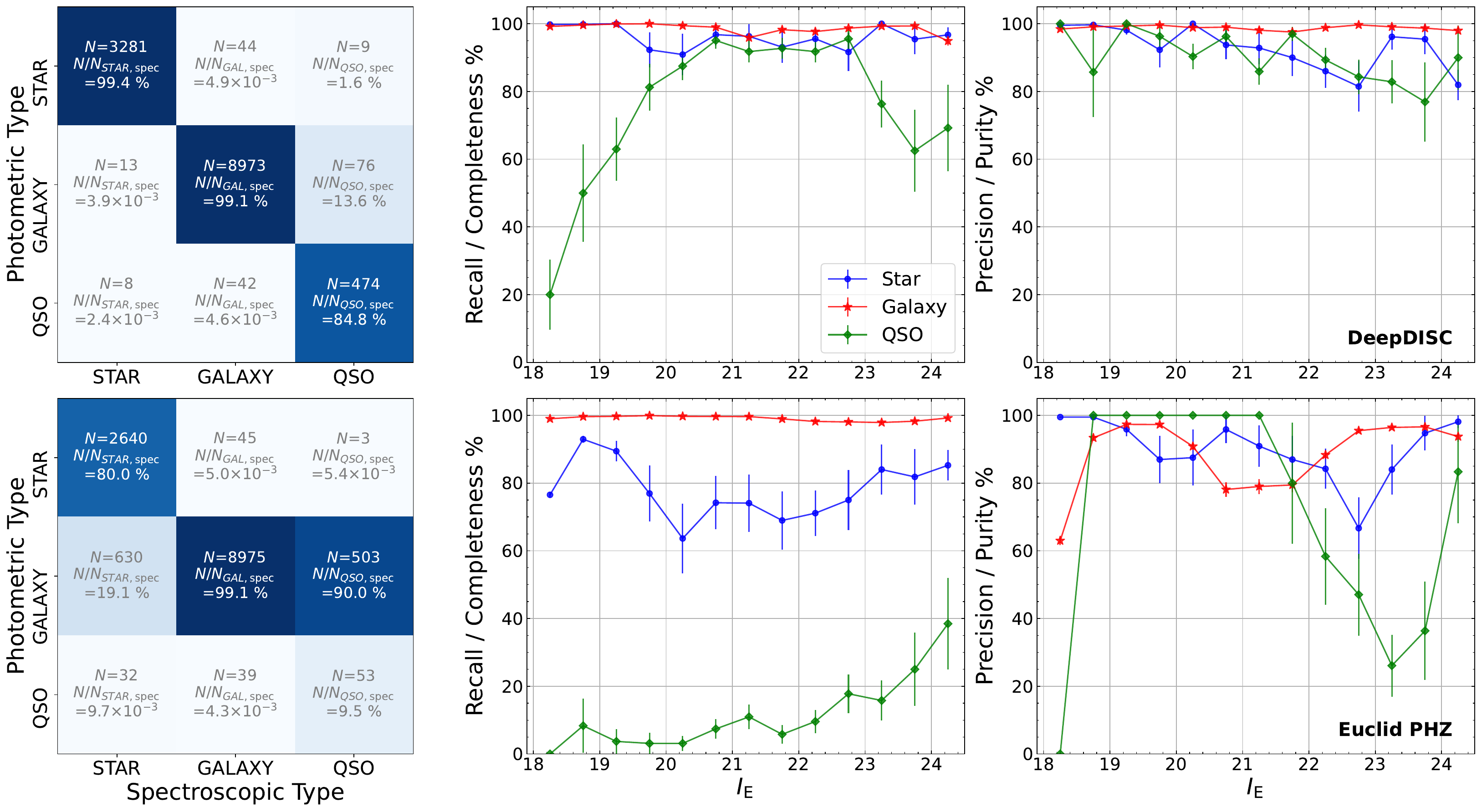}
 \caption{Comparison of source classification performance among matched sources from DeepDISC Model 1, the Euclid PHZ catalog \citepalias{Tucci25}, and DESI DR1. The analysis is further restricted to sources with a single classification in the Euclid PHZ catalog, corresponding to classifications with dominant probabilities in Euclid. The upper and lower panels show the results for DeepDISC and the Euclid PHZ catalog, respectively. The left column presents the confusion matrices of the classifications; the middle column shows recall (completeness) as a function of $\Ie$ band magnitude; and the right column shows precision as a function of $\Ie$ band magnitude.}
\label{fig:class_ephz}
\end{figure*}

Figure \ref{fig:class_ephz} compares the DeepDISC classifications (upper panels) with those from the Euclid PHZ catalog (lower panels). The two approaches deliver comparable performance for galaxy and star classifications. The completeness of star classification from DeepDISC is slightly larger than Euclid PHZ. Although the star classification precision seems different at the faint end, the two catalogs deliver comparable results within $\sim 1\sigma$. For the quasar classification, however, DeepDISC exhibits better performance than Euclid PHZ in both completeness and precision. \citetalias{Tucci25} stated that they produce a high-purity star sample, a highly-complete galaxy sample, but there was some difficulty to accurately separate quasars with galaxies, consistent with the results shown in the lower panels in Figure \ref{fig:class_ephz}. 

\begin{figure*}
\centering
\includegraphics[width=\textwidth]{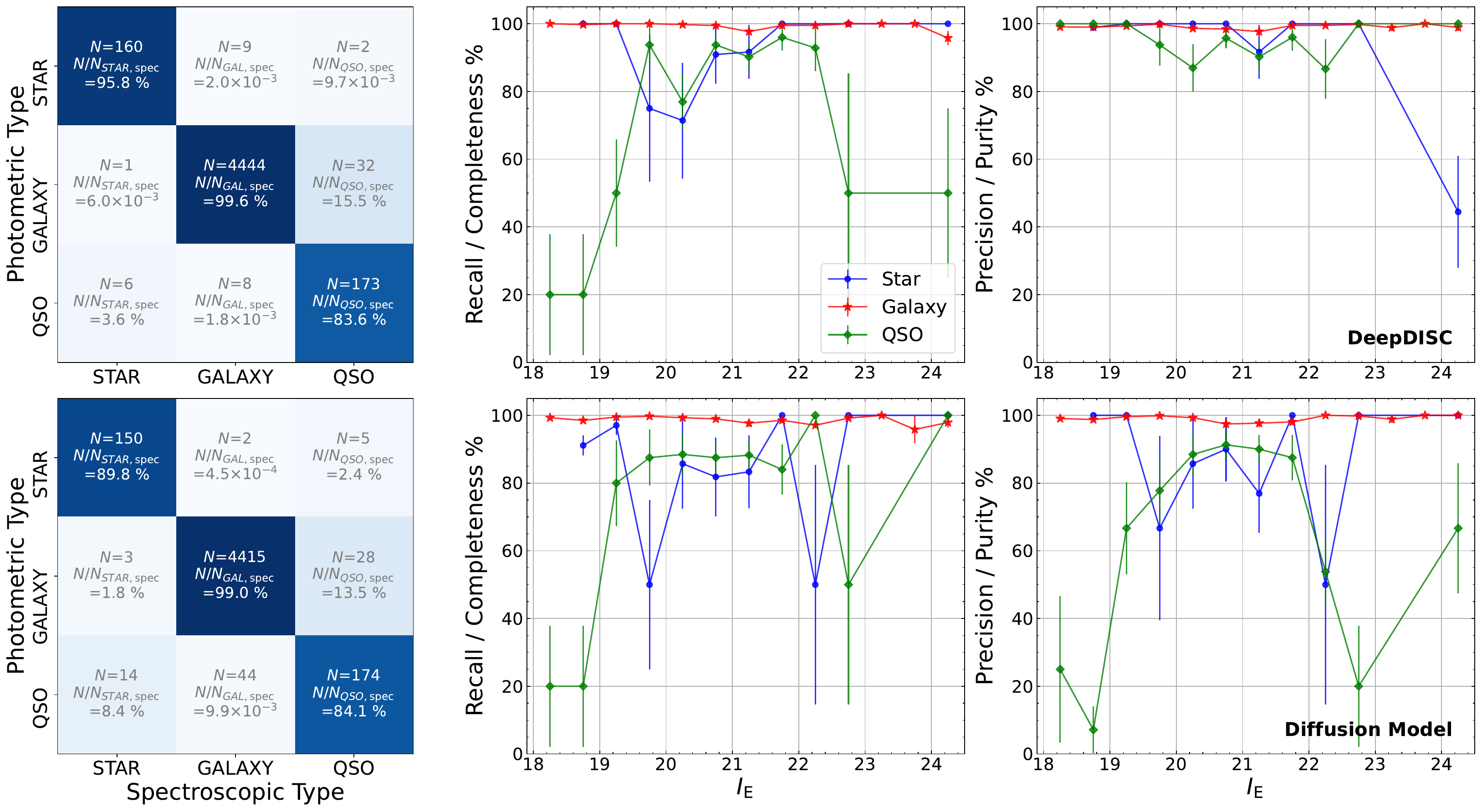}
 \caption{ Same as Figure \ref{fig:class_ephz}, but for matched sources among DeepDISC Model 1, the diffusion model predictions \citepalias{Stevens25}, and DESI DR1.}
\label{fig:class_df}
\end{figure*}

The comparison between DeepDISC and Diffusion Model classifications is displayed in Figure \ref{fig:class_df}. The sizes of the comparison samples are considerably smaller because \citetalias{Stevens25} applied stricter quality cuts than in this work to ensure that objects have sufficient SNR and are detected in the $\Ie$ band \citepalias[see details in][]{Stevens25}. Overall the two methods have comparable classification performance. The galaxy classifications are equally good for both methods. The star comparison samples are small because \citetalias{Stevens25} requires \texttt{det\_quality\_flag = 0}, which excludes many stars in the Euclid MER catalog, as many bright stars have non-zero \texttt{det\_quality\_flag}s due to flux saturation, surrounding structures (e.g., spikes), or blending. If we focus on the matched star sample across DESI, DeepDISC and Diffusion Model catalogs in the test set, the DeepDISC star classification has slightly better completeness and precision. Additionally, the completeness of quasar classification is comparable between DeepDISC and Diffusion Model, but the precision of DeepDISC quasar classification is better. The Diffusion Model predicted quasars have more fractional contamination from stars and galaxies. Furthermore, DeepDISC automatically separates quasars and stars, while the Diffusion Model approach requires additional steps for the star-quasar separation. 

\subsubsection{Redshift Comparison}
\label{sec:z_comp}

\begin{deluxetable*}{lccccccc}
\tablecaption{Photometric Redshift Comparison for Galaxies \label{tbl:z_gal_comp}}
\tablehead{
\colhead{Reference} &
\colhead{$N_{\rm Galaxy,DESI}$} &
\colhead{$e_{z,\mathrm{med,Model}}$} &
\colhead{$\sigma_{\rm IQR,Model}$} &
\colhead{$\eta_{\rm Model}$} &
\colhead{$e_{z,\mathrm{med,Ref}}$} &
\colhead{$\sigma_{\rm IQR,Ref}$} &
\colhead{$\eta_{\rm Ref}$}\\
\colhead{(1)} &
\colhead{(2)} &
\colhead{(3)} &
\colhead{(4)} &
\colhead{(5)} &
\colhead{(6)} &
\colhead{(7)} &
\colhead{(8)}
}
\startdata
This work & 9,579 & $2.0^{+0.5}_{-0.5}\times10^{-3}$ & $3.93^{+0.05}_{-0.05}\times10^{-2}$ & $5.77^{+0.24}_{-0.24}\%$  & \nodata & \nodata & \nodata\\ 
(Model 2) \\
Euclid PHZ galaxy & 8,967 & $2.0^{+0.5}_{-0.5}\times10^{-3}$ & $3.92^{+0.05}_{-0.05}\times10^{-2}$ & $5.04^{+0.26}_{-0.23}\%$  & $-2.40^{+0.05}_{-0.05}\times10^{-2}$ & $3.86^{+0.04}_{-0.04}\times10^{-2}$ & $6.06^{+0.25}_{-0.23}\%$ \\
(MT25; single flag) \\
AstroPT           & 413 & $5.3^{+2.1}_{-2.1}\times10^{-3}$ & $3.70^{+0.22}_{-0.23}\times10^{-2}$ & $3.06^{+1.18}_{-0.94}\%$  & $2.1^{+2.2}_{-2.2}\times10^{-3}$ &$4.40^{+0.21}_{-0.25}\times10^{-2}$ & $2.59^{+0.94}_{-0.71}\%$ \\
(MS25) 
\enddata
\tablecomments{Number of objects and associated statistics used for galaxy photo-$z$ comparison. Column (2) lists the sample sizes for galaxy photo-$z$ comparisons. In the first row, objects are required to be classified as galaxies by both DeepDISC and DESI in the test set. In subsequent rows, objects are additionally cross-matched with the corresponding catalogs. The Euclid PHZ sample includes only sources with a single galaxy classification flag. Columns (3)–(5) report the median bias relative to DESI spec-$z$, scatter, and outlier fraction for DeepDISC (Model 2), while columns (6)–(8) give the same metrics for the reference methods. Definitions of these statistics are provided in Section \ref{sec:z}.}
\end{deluxetable*}

\begin{figure*}
\centering
\includegraphics[width=\textwidth]{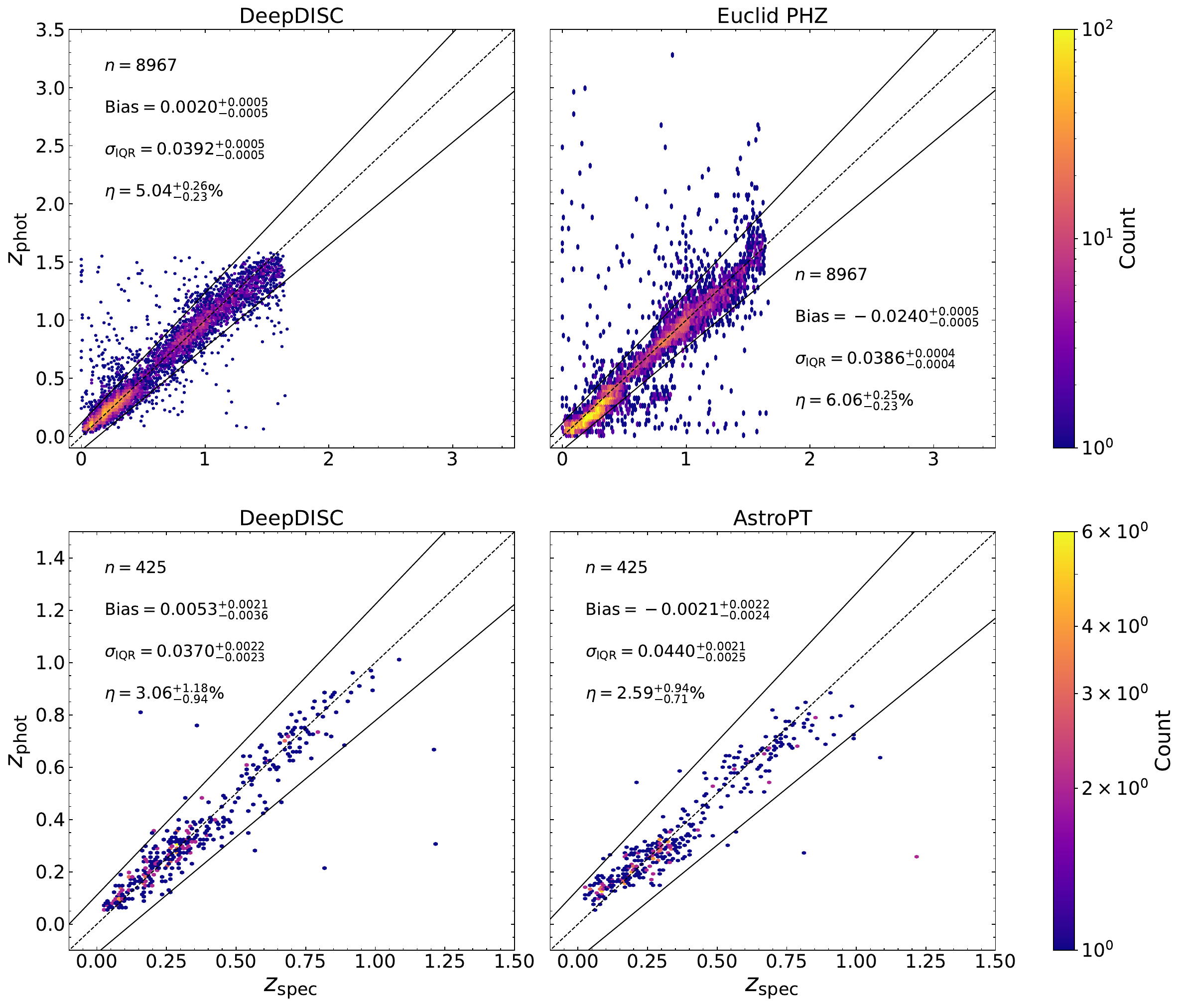}
 \caption{Comparison of galaxy photo-$z$ estimates from DeepDISC with those from the Euclid PHZ catalog \citepalias[upper][]{Tucci25} and the AstroPT catalog \citepalias[lower][]{Siudek2025}. The left column shows the DeepDISC results, while the right column shows the Euclid PHZ (upper panel) and AstroPT (lower panel) results. Sources included in each comparison are required to have DESI spec-$z$s as well as photo-$z$s from DeepDISC and Euclid PHZ/AstroPT. 
 The color in each hexagonal bin indicates the number of sources in logarithmic scale, as shown by the color bar. The median bias relative to $\zspec$, scatter $\sigma_{\rm IQR}$, and outlier fraction $\eta$ (defined in Section \ref{sec:z}) for each sample are also shown in each panel. The 1:1 relation and the outlier boundaries at different redshifts are indicated by dashed and solid lines, respectively.}
\label{fig:z_gal_comp}
\end{figure*}

We compile photo-$z$s from different studies and compare them with the DeepDISC inference results on the test set, for which spectroscopic redshifts are available. The Euclid PHZ catalog \citepalias{Tucci25} and \citet{Siudek2025} (\citetalias{Siudek2025}) are used for the galaxy photo-$z$ comparison, while the Euclid PHZ catalog and \citet{Roster2025} (\citetalias{Roster2025}) are used for the quasar photo-$z$ comparison. For both the galaxy and quasar photo-$z$ comparisons, we require objects to be consistently classified as galaxies or quasars by both DESI and the corresponding catalog in the test, in order to avoid potential degeneracies between classification and redshift estimation. For the galaxy photo-$z$ comparison, we further require a strict one-to-one comparison, ensuring that each object has photo-$z$ measurements from DESI, DeepDISC, and Euclid PHZ or \citetalias{Siudek2025}. For the quasar photo-$z$ comparison, we relax this requirement and cross-match each photo-$z$ sample separately with DESI to increase the sample size for each comparison sample, given the limited number of quasars in each catalog. The number of sources and corresponding statistics used in the comparison are summarized in Tables~\ref{tbl:z_gal_comp} and \ref{tbl:z_qso_comp} for galaxies and quasars, respectively.


Figure \ref{fig:z_gal_comp} presents the comparison of DeepDISC galaxy photo-$z$s with those reported by \citetalias{Tucci25} and \citetalias{Siudek2025}, respectively. The Euclid PHZ catalog calculates the photo-$z$s through the template-fitting tool Phosphoros \citep{Tucci25}. It fits the observed flux with galaxy-only SED templates and a set of parameters, including redshift, intrinsic reddening curve and intrinsic attenuation $E(B-V)$. The statistics for photo-$z$ comparison defined in Section \ref{sec:z} are also computed for each sample. The median bias $e_{z,{\rm med}}$, scatter $\sigma_{\rm IQR}$, and outlier fraction $\eta$ for sources in this comparison predicted by DeepDISC are $2.0^{+0.5}_{-0.5}\times10^{-3}$, $3.92^{+0.05}_{-0.05}\times10^{-2}$, and $5.04^{+0.26}_{-0.23}\%$, respectively, while those given by the Euclid PHZ catalog are $-2.4^{+0.05}_{-0.05}\times10^{-2}$, $3.86^{+0.04}_{-0.04}\times10^{-2}$, and $6.06^{+0.25}_{-0.23}\%$. photo-$z$ predictions are generally consistent with DESI spec-$z$s with small biases reported above. However, the Euclid photo-$z$s exhibits a significantly larger offset, exceeding that of DeepDISC by approximately one order of magnitude. The scatter of DeepDISC and Euclid PHZ galaxy photo-$z$s remains consistent within $1\sigma$, but DeepDISC shows a somewhat lower outlier fraction. Overall, DeepDISC delivers a smaller bias with respect to DESI spec-$z$s, comparable scatter, and a reduced outlier fraction compared to the Euclid PHZ catalog. 

The lower panels in Figure \ref{fig:z_gal_comp} display the comparison between DeepDISC photo-$z$s with \citetalias{Siudek2025} results for galaxies. \citetalias{Siudek2025} trained a multi-modal foundation model using Euclid Q1 optical and near-infrared image cutouts together with SED information (VIS+NISP+SED) via AstroPT, and subsequently fine-tuned the model for multiple downstream tasks, including photo-$z$ estimation for selected galaxies. On the contrary, the DeepDISC models are image-based and trained with UNIONS+VIS+NISP images only. For convenience, we hereafter refer to this model as the AstroPT model. Their analysis selects sources with segmentation areas larger than 800 pixels and $\He < 22.5$, which predominantly correspond to bright, extended galaxies. We therefore caution that this comparison is not representative of the full Euclid source population; both methods tend to perform well on bright, extended objects, so the observed parity does not imply equivalent performance at fainter magnitudes or for compact sources. More broadly, foundation models such as AstroPT are optimized for general-purpose representation learning and reconstruction objectives rather than for task-specific performance, and their fine-tuned downstream predictions do not necessarily outperform dedicated supervised models. As demonstrated by \citet{Ting2025}, traditional supervised learning remains preferred for task-specific inference (including photometric redshift estimation) when the training sample is sufficiently large, because the supervised model can directly optimize the task loss rather than relying on representations learned under a different objective. The DESI DR1 training sample used here, with $\sim$40,000 labeled sources in EDF-N, is large enough to place DeepDISC in this favorable regime, explaining the comparable or better performance of our purpose-built supervised model relative to the fine-tuned AstroPT foundation model on the overlapping bright galaxy sample. 



As shown in the lower panels of Figure~\ref{fig:z_gal_comp} and Table~\ref{tbl:z_gal_comp}, DeepDISC and AstroPT achieve comparable performance for galaxy photo-$z$. The median bias $e_{z,{\rm med}}$, scatter $\sigma_{\rm IQR}$, and outlier fraction $\eta$ for sources in this comparison predicted by DeepDISC are $5.3^{+2.1}_{-3.6}\times10^{-3}$, $3.70^{+0.22}_{-0.23}\times10^{-2}$, and $3.06^{+1.18}_{-0.94}\%$, respectively, while those given by AstroPT is $-2.1^{+2.2}_{-2.4}\times10^{-3}$, $4.40^{+0.21}_{-0.25}\times10^{-2}$, and $2.59^{+0.94}_{-0.71}\%$. The median biases relative to DESI spec-$z$ are of the same order. The scatters agree within $2\sigma$, and the outlier fractions are consistent within $1\sigma$. Notably, DeepDISC does not require a pre-classified galaxy sample for training and can be applied to a larger and more diverse sample with reliable photo-$z$.

We also compare the quasar photo-$z$ performance for DeepDISC, Euclid PHZ catalog, and \citetalias{Roster2025}. As mentioned above, the Euclid PHZ catalog adopts the template-fitting Phosphoros tool and fits the observed flux only with galaxy SED, which is not optimized for quasar $\zphot$ measurements. In contrast, \citetalias{Roster2025} identify extragalactic X-ray counterparts as quasars (hereafter the X-ray CTP catalog) and estimate their photo-$z$s using PICZL \citep{Roster2024}, an image-based deep-learning method. The extragalactic X-ray counterparts in their catalog are cross-matched with Legacy Survey Data Release 10 \citep[LS10;][]{Dey2019} by NWAY, a Bayesian algorithm for cross-matching \citep{NWAY}. \citetalias{Roster2025} measures photo-$z$s for matched objects, utilizing LS10 calibrated, de-reddened $griz$ images and $W1-W2$ flux of WISE photometry, with additional pre-processing and feature construction as described in \citet{Roster2024}. 

\begin{figure*}
\centering
\includegraphics[width=\textwidth]{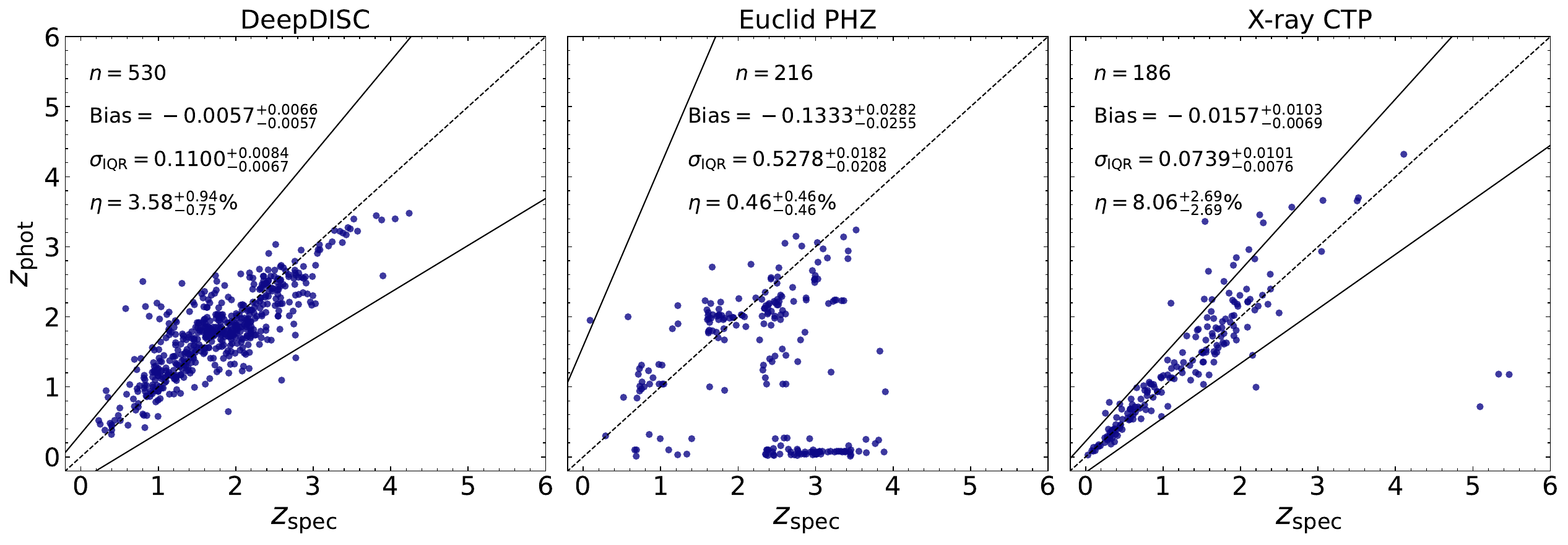}
 \caption{Comparison of quasar photo-$z$ estimates from DeepDISC (left) with those from the Euclid PHZ catalog \citepalias[middle][]{Tucci25} and the X-ray CTP catalog \citepalias[right][]{Roster2025}. Sources included in each comparison are required to have DESI spectroscopic counterparts. The median bias relative to $\zspec$, scatter $\sigma_{\rm IQR}$, and outlier fraction $\eta$ (defined in Section \ref{sec:z}) for each sample are shown in each panel. The 1:1 relation and the outlier boundaries at different redshifts are indicated by dashed and solid lines, respectively.}
\label{fig:z_qso_comp}
\end{figure*}

The comparison of QSO photo-$z$s is shown in Figure \ref{fig:z_qso_comp} and Table \ref{tbl:z_qso_comp}. The comparison is for all objects matched with the DESI quasar sample, in order to retain a large sample in each catalog. Although this comparison for quasar photo-$z$s is not a one-to-one comparison, the matched DESI quasars in all three catalogs have similar redshift and magnitude distributions, ensuring a fair comparison. The left and middle panels of Figure \ref{fig:z_qso_comp} compare the photo-$z$s predicted by DeepDISC and Euclid PHZ catalog, respectively, with DESI spec-$z$s. DeepDISC quasar photo-$z$s show good agreement with DESI spec-$z$s with slightly underestimation of $e_{z,{\rm med}} = -5.7^{+6.6}_{-5.7}\times10^{-3}$, of which the absolute value is about of 2 order of magnitude lower than that for Euclid PHZ catalog, $-1.333^{+0.282}_{-0.255}\times10^{-1}$. The scatter of DeepDISC photo-$z$ is $1.100^{+0.084}_{-0.067}\times10^{-1}$, 5 times smaller than that for Euclid PHZ catalog of $5.278^{+0.182}_{-0.208} \times 10^{-1}$. The outlier fraction for DeepDISC and Euclid PHZ catalog are $3.58^{+0.94}_{-0.75}\%$ and $0.46^{+0.46}_{-0.46}\%$, respectively. The Euclid PHZ catalog has a notably smaller outlier fraction because of the extremely larger scatter. The right panel display redshifts from objects from the X-ray CTP catalog matched with DESI. The median bias $e_{z,{\rm med}}$, scatter $\sigma_{\rm IQR}$ and outlier fraction $\eta$ is $-1.57^{+1.03}_{-0.69}\times10^{-2}$, $7.39^{+1.01}_{-0.76}\times10^{-2}$, and $8.06^{+2.69}_{-2.69}\%$, respectively. The X-ray CTP also slightly underestimates the quasar redshifts. And its absolute median bias is larger than that of the DeepDISC catalog by a factor of 2.8, although the discrepancy remains consistent within $1\sigma$. Although DeepDISC exhibits a larger scatter by a factor of 1.5, the scatter values of the two methods are still comparable within $2\sigma$. Additionally, the outlier fraction is also consistent within $2\sigma$, and the modestly smaller fraction for DeepDISC is mainly due to the larger scatter.  

In summary, DeepDISC provides accurate quasar photo-$z$ estimates, outperforming the Euclid PHZ catalog and achieving performance comparable to the X-ray CTP catalog. Moreover, DeepDISC also offers detection and classification for all Euclid sources, whereas X-ray CTP catalog using PICZL relies on a pre-selected type of objects. DeepDISC can therefore provide a larger photo-$z$ sample for both galaxies and quasars in Euclid fields.



\begin{deluxetable*}{lccccc}
\tablecaption{Photometric Redshift Comparison for QSOs\label{tbl:z_qso_comp}}
\tablehead{
\colhead{Reference} &
\colhead{$N_{\rm QSO,DESI}$} &
\colhead{$\mathrm{Med}\, e_{z}$} &
\colhead{$\sigma_{\rm IQR}$} &
\colhead{$\eta$} &\\
\colhead{(1)} &
\colhead{(2)} &
\colhead{(3)} &
\colhead{(4)} &
\colhead{(5)} &
}
\startdata
This work & 530 & $-5.7^{+6.6}_{-5.7}\times10^{-3}$ & $1.100^{+0.084}_{-0.067}\times10^{-1}$ & $3.58^{+0.94}_{-0.75}\%$ \\
(Model 3) \\
Euclid PHZ quasar & 216 & $-1.333^{+0.282}_{-0.255}\times10^{-1}$ & $5.278^{+0.182}_{-0.208}\times10^{-1}$ & $0.46^{+0.46}_{-0.46}
\%$ \\
(MT25; single flag) \\
X-ray CTP  & 186 & $-1.57^{+1.03}_{-0.69}\times10^{-2}$ & $7.39^{+1.01}_{-0.76}\times10^{-2}$ & $8.06^{+2.69}_{-2.69}\%$ \\
(WR25)
\enddata
\tablecomments{Number of objects and associated statistics used for quasar photo-$z$ comparison. Column (2) lists the sample sizes. Objects are required to be classified as quasars by both DESI and the corresponding method in the test set. The Euclid PHZ sample includes only sources with a single quasar classification flag. The comparison is not restricted to one-to-one matches due to the limited quasar sample size. Columns (3)–(5) report the median bias relative to DESI spec-$z$, scatter, and outlier fraction for DeepDISC, Euclid PHZ, and the X-ray CTP catalog, respectively. Definitions of these statistics are given in Section~\ref{sec:z}.}
\end{deluxetable*}

\subsection{Verification with Euclid Q1 spectra}

Compared with several earlier studies, the main improvement in terms of source classification and photometric redshift estimation with our DeepDISC approach is for broad-line quasars. We visually inspected the Euclid grism spectra for EDF-N sources classified as quasars by DeepDISC and have photometric redshifts within $1<z_{\rm phot}<2$. Unfortunately, the current Q1 release of grism spectra does not have sufficient quality for most sources for a reliable evaluation of the classification and redshift estimation. Upon visual inspection of random sources, there are many clean examples where the DeepDISC classification and photometric redshift are correct, though most of the spectra suffer from quality issues that prevent robust typing \citep{Euclid:SPE}.





\subsection{Caveats of the DeepDISC catalog}\label{sec:caveats}

\begin{figure*}
\centering
\includegraphics[width=0.9\textwidth]{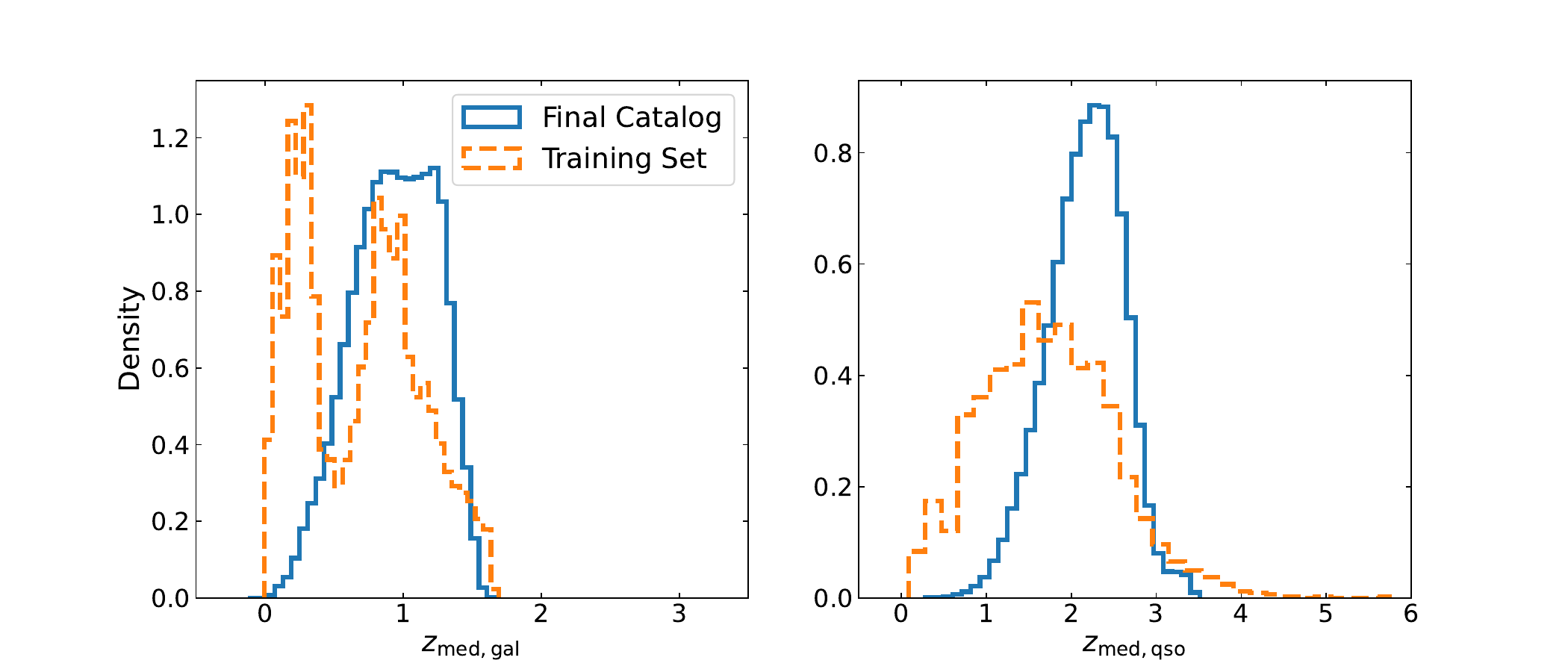}
\caption{Redshift distributions of galaxies (left) and quasars (right) in the final DeepDISC catalog (Section~\ref{sec:cat}) compared with those of the training set. Blue solid lines denote the DeepDISC distributions, while orange dashed lines denote the DESI training-set distributions.}
\label{fig:final_z}
\end{figure*}

In Sections \ref{sec:res} and \ref{sec:comp}, we demonstrate that DeepDISC achieves good performance in detection, classification, and redshift estimation. However, we outline several important caveats and limitations of the DeepDISC catalog upon application. These caveats are fully expected for deep-learning models, and are produced inherently by the quality of the training data set. 

First, the DeepDISC detections are somewhat problematic for faint objects, {especially objects with magnitudes fainter than 24 in all bands}, {where photon noise becomes significant in model forecast}. As described in Section \ref{sec:det}, visual inspection indicates that detections for source brighter than 24 magnitude {in at least one band} would be mostly reliable. A related consequence is that for multiple DeepDISC detections at overlapping positions, if they are faint in all bands, one of the detection would be simply caused by noise; however, if one of them are faint in $\Ie$ band, but bright in other bands (e.g., $\Ye$, $\Je$, and $\He$), these multiple detections may be two distinct objects or structures with different colors at the same position. Indeed, some examples we showed in Figure \ref{fig:multi_deblend} demonstrate that DeepDISC is able to deblend close objects using multi-band information simultaneously. For a clean sample, we suggest retaining only one source among multiple detections at nearly the same position when the companion objects are faint in all Euclid bands. To confirm cases of true close companions, visual inspection of multi-band cutouts is recommended.

In addition, the positions and fluxes for the detected sources compiled in Table~\ref{tbl:catalog} are calculated following simple recipes. For accurate astrometry (e.g., better than $\sim 0\farcs3$) and photometry, additional processing is necessary using the provided source segmentation box. Alternatively, more accurate outputs of source positions and fluxes directly from the deep learning model are also possible, if these labels are added in the training. Future DeepDISC implementations will include such improvements (Ejjagiri et al., in prep).

Second, model classification and redshift estimation are limited to the parameter ranges in the training set (DESI DR1 catalogs). While the DESI star, galaxy and QSO samples cover broad magnitude ranges (Figure \ref{fig:mag_dist}), DeepDISC source classification beyond these magnitude ranges may be unreliable. In addition, the DeepDISC photometric redshifts cannot go beyond the redshift ranges of galaxies and QSOs of their corresponding sets in the DESI training sample (Figure \ref{fig:z_dist}). For QSOs, the DESI redshift sample covers a broad redshift range of $0\lesssim z\lesssim 5$, although there are fewer objects at high-$z$ than those at lower redshift, so this is not a severe problem. The DESI galaxy redshift sample only covers to $z\approx 1.8$, meaning the DeepDISC catalog will not report photo-$z$s beyond the DESI galaxy redshift range. Fortunately, for a flux-limited sample at the Euclid depth, most galaxies will peak around $z\sim 1-2$ (Cosmic Noon), and only a small fraction of $z\gtrsim 2$ galaxies in EDF-N will be assigned underestimated redshifts by DeepDICS (but these galaxies are still detected by DeepDISC). Figure~\ref{fig:final_z} shows the photo-$z$ distributions for galaxies and QSOs in the DeepDISC catalog. These photo-$z$ distributions do not follow those of the training sample, ensuring that DeepDISC learns real features rather than randomly sampling from the redshift distributions of the training set. Importantly, DeepDISC classified many of the low-$z$ quasars as galaxies (as discussed in \S\ref{sec:class}), leading to a different redshift distribution than DESI quasars.


These important caveats underscore the need for better training samples that cover the full range of parameters of Euclid sources. The EDF-N is being targeted by Euclid grism repeatedly \citep{Euclid}, as well as by additional spectroscopic programs with, e.g., Subaru-PFS \citep{Takada2014} and JWST \citep{nexus}. Improved redshift samples within EDF-N over the next few years can significantly improve the performance of DeepDISC on the full Euclid data. The upcoming Euclid DR1 will cover a substantially larger sky area, providing access to significantly larger existing spectroscopic training samples from overlapping surveys, directly addressing the labeled-data limitations that motivate our staged training strategy and restrict our current galaxy photo-$z$ coverage to $z\leq1.8$. In addition, ongoing JWST programs such as NEXUS \citep{nexus,Zhuang2024,Zhuang2026} and dedicated spectroscopic training sample efforts including C3R2 \citep{Masters2017C3R2,Stanford2021} and 4C3R2 \citep{Guglielmo2020,Gruen2023} will provide spectroscopic redshifts at $z\gtrsim2$ and systematic coverage of underrepresented regions of color-redshift space, enabling DeepDISC galaxy photo-$z$ estimation to extend to higher redshifts and reducing out-of-distribution bias in the predicted photo-$z$ PDFs. Together, these expanded training resources will enable a fully joint training regime for detection, deblending, classification, and photo-$z$ estimation, unlocking the full mutual benefit of these tasks within the DeepDISC framework.

Several methodological directions can address these limitations in future work as well. First, the labeled-data bottleneck can be alleviated by semi-supervised approaches that combine supervised learning on spectroscopically confirmed objects with self-supervised learning on the much larger pool of unlabeled Euclid photometric sources. \citet{Khederlarian2026} demonstrate this using {\it HST}/CANDELS \citep{Koekemoer2011} imaging out to $z\sim3$, showing that their semi-supervised model PITA outperforms fully supervised and classical methods even when the spectroscopic training set is significantly reduced, which is a regime directly analogous to the sparse DESI coverage at high redshift in EDF-N. Second, systematic training sample augmentation using techniques developed for wide-field surveys, including empirical reweighting of underrepresented regions of color-redshift space \citep{Moskowitz2024,Zhang2025b} and generative augmentation of source populations with few spectroscopic labels \citep{Moran2026}, can extend reliable photo-$z$ coverage to redshifts and source types poorly sampled by DESI DR1. Third, joint processing of Euclid and Rubin/LSST \citep{Ivezic2019} imaging will substantially improve photo-$z$ accuracy \citep{Chary2019}: the combination of Rubin's broad optical coverage across six bands with Euclid's near-infrared imaging provides both greater spectral baseline and finer sampling of the spectral energy distribution than either survey alone, breaking color-redshift degeneracies that arise when optical or near-infrared data are available in isolation, and enabling more accurate photo-$z$ estimation across a wide redshift range. Together these advances will position DeepDISC to deliver photo-$z$ PDFs meeting the stringent requirements of Euclid's weak lensing and galaxy clustering science cases. 




\section{Conclusions}\label{sec:con}

In this work, we utilize a deep learning framework, DeepDISC, to train models for source detection, classification and photo-$z$ estimation on Euclid Q1 images in EDF-N. Three models are trained: (1) Model 1 for detection and classification using Euclid Q1 segmentation maps and DESI DR1 spectroscopic classifications (stars, galaxies and quasars) as ground truth; (2) Model 2 for galaxy redshifts using DESI DR1 galaxy spec-$z$ as ground truth; (3) Model 3 for quasar redshifts using DESI DR1 quasar spec-$z$ as ground truth. We apply the trained model to all EDF-N images and present a catalog of 13,355,031 sources with source coordinates, classifications as well as photometric redshifts for galaxies and quasars. 

We compare the performance of our DeepDISC method with several earlier Euclid studies that presented source catalogs, classifications, and/or photometric redshifts \citep{Roster2025,Siudek2025,Stevens25,Tucci25}. Overall, DeepDISC performs equally well for galaxies and stars, with improved performance for quasar classification and photo-$z$s. 


\begin{itemize}
    \item  The detection completeness (recall) and purity (precision) for all objects by DeepDISC Model 1 are ${\sim}93\%$ and ${\sim}80\%$, respectively, if using the Euclid MER source catalog as the benchmark. The true purity is even higher ($\sim 90\%$) when using a JWST source catalog that covers a small fraction of EDF-N as the benchmark. 
    
    \item The detection/classification module (Model 1) also identifies and recovers correct types for $99.2\%$, $99.0\%$, and $84.8\%$ DESI stars, galaxies, and quasars. The model has comparable classification performance as the Euclid PHZ \citepalias{Tucci25} for stars and galaxies, and improved performance for quasar classifications. It also has comparable classification performance as the diffusion model in \citetalias{Stevens25}, where the latter does not distinguish between stars and quasars.
    
    \item The galaxy photo-$z$s predicted by Model 2 are in good agreement with DESI DR1 spec-$z$s, exhibiting a median bias of $2.0\times10^{-3}$. The overall photo-$z$ performance is comparable to that of Euclid PHZ, but with a lower outlier fraction in the DeepDISC predictions. DeepDISC also achieves similar performance on galaxy photo-$z$s as AstroPT \citepalias{Siudek2025}, but without requiring pre-selected galaxy samples.
    
    \item The quasar photo-$z$s predicted by Model 3 generally agree with DESI DR1 spec-$z$s, with a median bias of $-5.7\times10^{-3}$. The overall photo-$z$ performance is better than that of Euclid PHZ, which is designed primarily for galaxies. Model 3 achieves comparable photo-$z$ accuracy as the X-ray CTP catalog provided by PICZL \citepalias{Roster2025}, although DeepDISC results have larger scatter. Notably, DeepDISC does not require quasars to be pre-selected, enabling the construction of a larger and more complete sample. 
\end{itemize}



The compiled DeepDISC source catalog of $\sim$13 million objects in EDF-N provides an independent, image-based data product complementary to the Euclid PHZ pipeline, with demonstrated advantages in quasar classification and photometric redshift estimation. This catalog enables a broad range of immediate science applications, including construction of cleaner quasar samples for clustering and reionization studies, photometric galaxy samples for weak lensing and galaxy evolution analyses, and cross-matching with multi-wavelength datasets targeting EDF-N. Looking forward, the EDF-N will be observed repeatedly by Euclid grism spectroscopy and targeted by dedicated programs including Subaru-PFS and JWST NEXUS, progressively building the spectroscopic training sample needed to push DeepDISC photo-$z$ coverage beyond the current $z {\approx}1.8$ ceiling imposed by DESI DR1 and to improve quasar completeness at low redshift where host galaxy contamination currently limits performance. On the methodological side, semi-supervised learning \citep{Khederlarian2026}, training sample augmentation techniques \citep{Moskowitz2024,Zhang2025b,Moran2026}, and joint Rubin+Euclid processing will collectively address the labeled-data bottleneck that motivates our current staged training strategy. As Euclid DR1 extends coverage to a substantially larger sky area with access to richer overlapping spectroscopic datasets, the DeepDISC framework developed here is directly scalable to full-survey deployment, ultimately delivering pixel-level deep learning based detection, deblending, classification, and probabilistic photo-$z$ estimation at the scale of the Euclid wide survey.

\section*{Acknowledgments}

We thank S. Luo at the National Center for Supercomputing Applications (NCSA) for helpful discussion and assistance with the GPU cluster used in this work.
This work is supported by NASA grant 80NSSC26K0333. G. Merz, S. Lin, and X. Liu also acknowledge support from the Illinois Campus Research Board Award RB25035, NSF grant AST-2308174, and NASA grant 80NSSC24K0219.
%
This work used Delta and DeltaAI at NCSA through allocation PHY240290 from the Advanced Cyberinfrastructure Coordination Ecosystem: Services \& Support (ACCESS) program, which is supported by U.S. National Science Foundation grants \#2138259, \#2138286, \#2138307, \#2137603, and \#2138296. 
Based on observations with the NASA/ESA/CSA James Webb Space Telescope obtained from the Barbara A. Mikulski Archive at the Space Telescope Science Institute, which is operated
by the Association of Universities for Research in Astronomy, Incorporated, under NASA contract NAS5-03127. Support for Program number JWST-GO-05105 (NEXUS) was provided through a grant from the STScI under NASA contract NAS5-03127. 

\bibliographystyle{aasjournal}

\bibliography{main}

\begin{appendix}

\section{Representative Examples of Unmatched Detections}
\label{appx1}
\begin{figure*}
\centering
\includegraphics[width=0.8\textwidth]{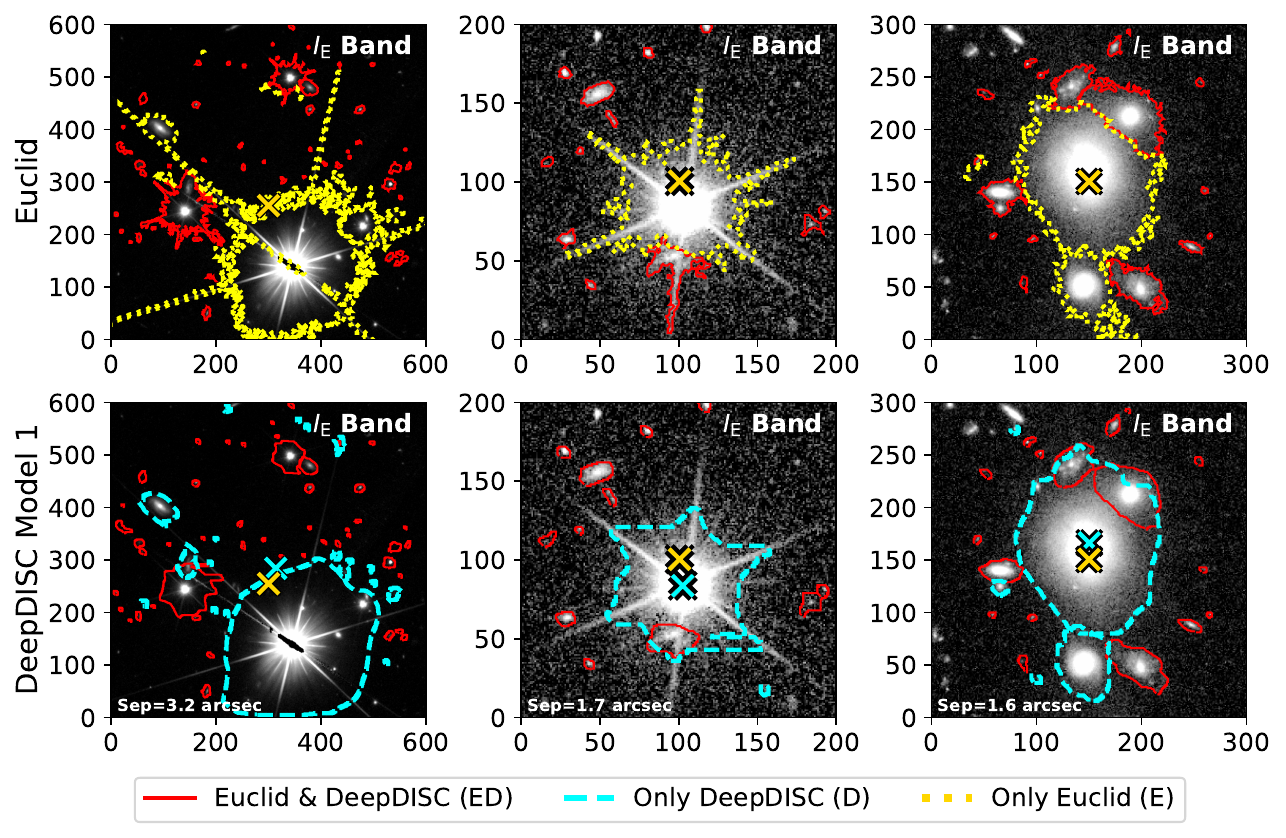}
\caption{Three examples of mismatches between Euclid (top panels) and DeepDISC (bottom panels) detections for bright stars or extended galaxies. The left panels show a star for which Euclid and DeepDISC predict similar centers, yet the offset still exceeds the cross-matching threshold. In both cases, the positions are defined as the centers of the bounding boxes, which can be biased by diffraction spikes. The middle panels show examples where the Euclid bounding-box center of a bright star is offset due to an overlapping detection along diffraction spikes. The right panels illustrate a case in which the Euclid detection of a galaxy is biased by nearby surrounding galaxies. These examples indicate that DeepDISC is able to detect most very bright or extended objects, although the predicted centroids may be offset relative to the Euclid catalog.}
\label{fig:ext}
\end{figure*}

\begin{figure*}
\centering
\includegraphics[width=0.8\textwidth]{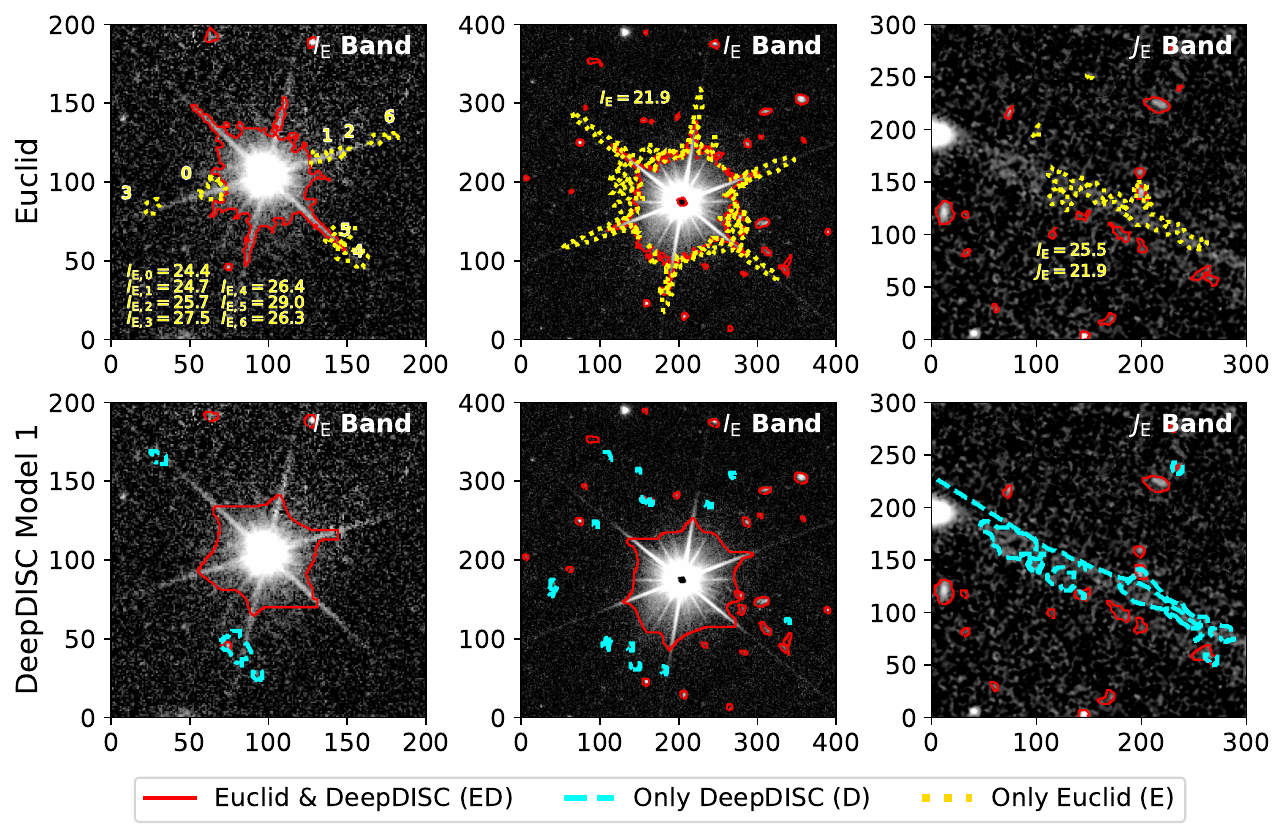}
\caption{Three examples of image artifacts affecting the detection. The top and bottom panels are for Euclid and DeepDISC detections, respectively. The left panels show suspicious Euclid detections along diffraction spikes of bright stars that are not detected by DeepDISC, which are usually very faint in $\Ie$ band ($\Ie>24$). The middle panels illustrate cases where Euclid detections include parts of the diffraction spikes as separate sources, while DeepDISC detects only the central star. Such spikes can become particularly bright for some stars. The right panels show other image artifacts that may be detected by both Euclid and DeepDISC. In this case, it may result from inaccurate background subtraction or imperfect stacking. These detections are irregular, leading to both Euclid-only (E) and DeepDISC-only (D) detections.}
\label{fig:artifact}
\end{figure*}

\begin{figure*}
\centering
\includegraphics[width=0.8\textwidth]{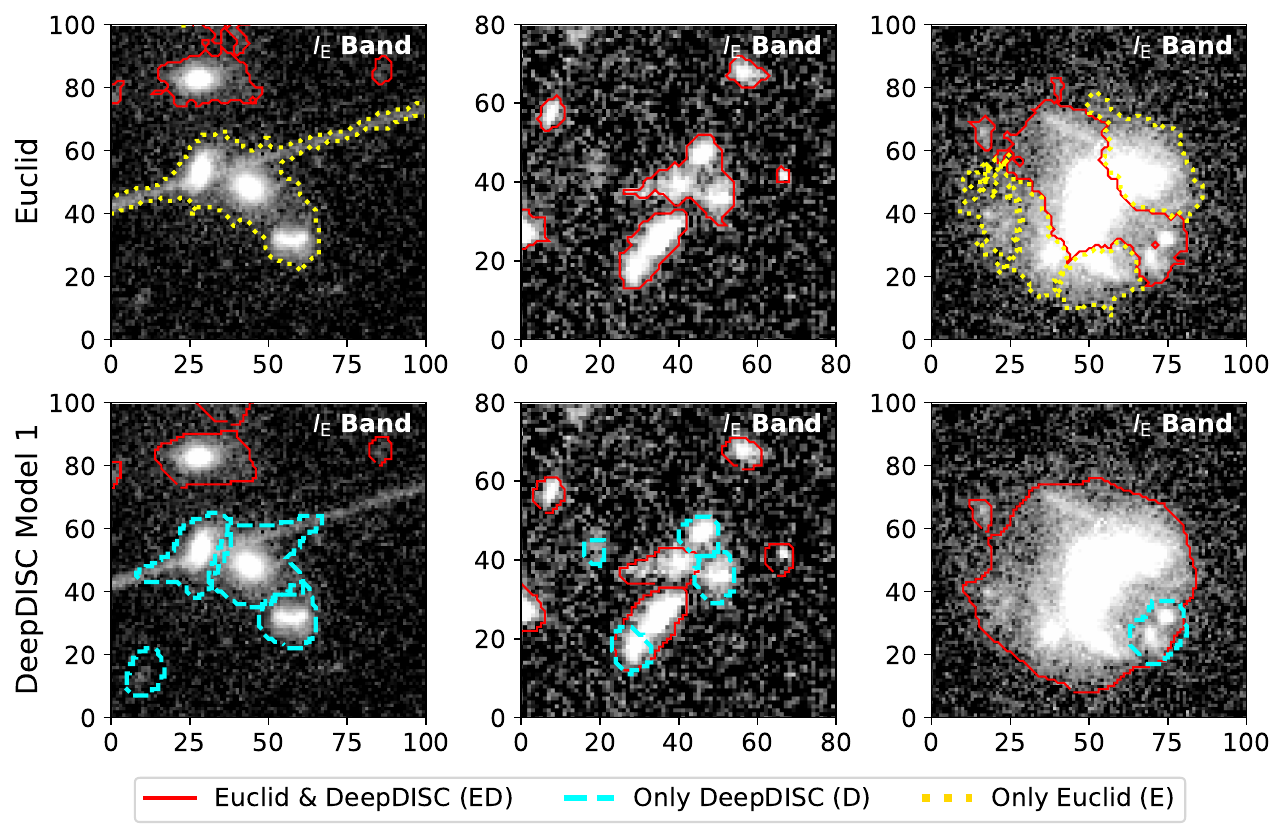}
\caption{Three examples of deblending by Euclid (top panels) and DeepDISC (bottom panels). The left panels illustrate DeepDISC separating sources embedded in a stellar diffraction spike. The middle panels present DeepDISC deblending of multiple closely spaced objects, while the right panels show the deblending of a complex system. These examples suggest that DeepDISC can more effectively separates close companions in the presence of artifacts from bright or extended structures. In contrast, deblending of multiple closely spaced objects remains challenging and appears more stochastic for both Euclid and DeepDISC.}
\label{fig:deblend}
\end{figure*}

\begin{figure*}
\centering
\includegraphics[width=0.8\textwidth]{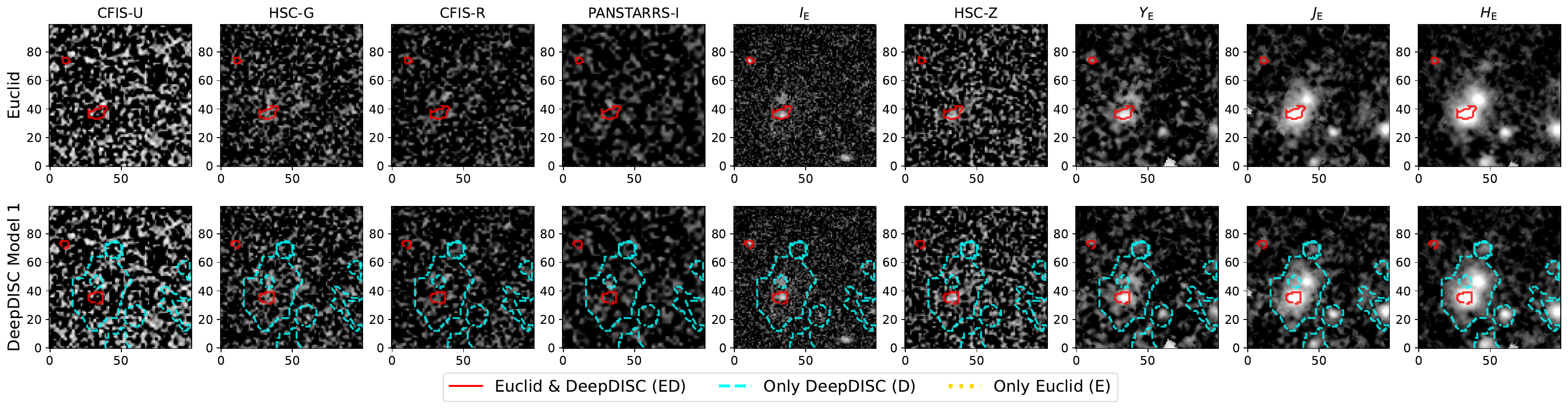}
\includegraphics[width=0.8\textwidth]{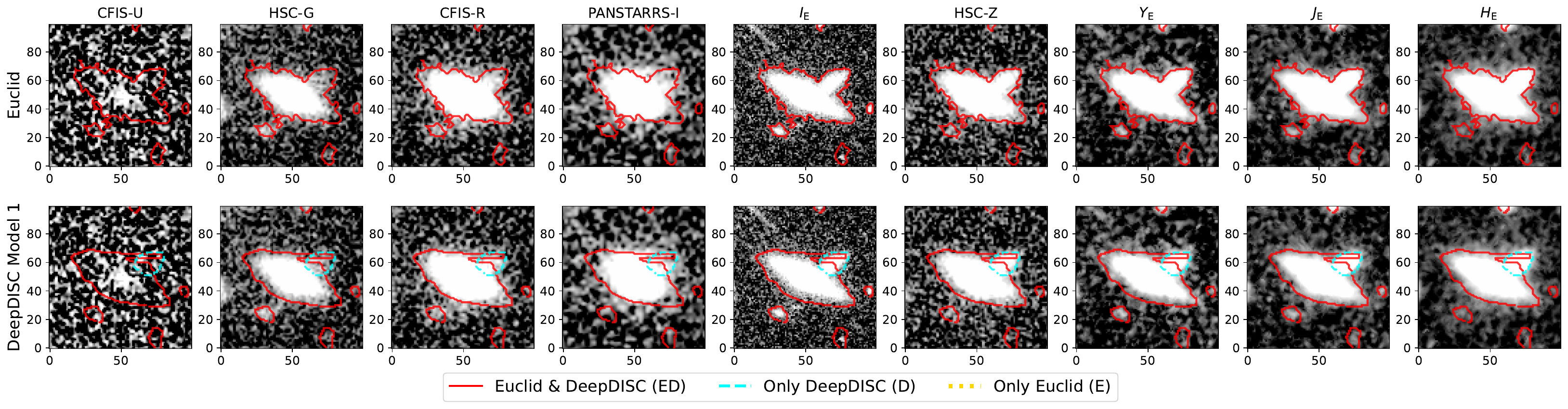}
\includegraphics[width=0.8\textwidth]{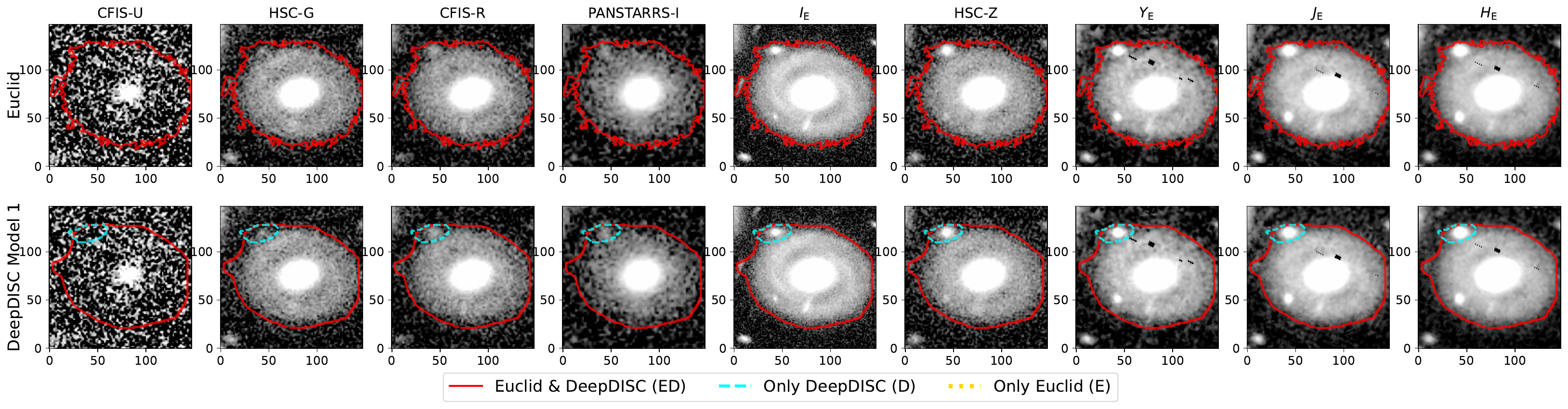}
\caption{Three examples illustrating how multi-band information improves DeepDISC detection and deblending. The top panels show a case where objects that are faint in the optical bands but bright in the near-infrared are missed by Euclid yet detected by DeepDISC. The middle and bottom panels present examples where DeepDISC successfully separates sources that are blended in Euclid MER detection by leveraging multi-band information. These examples demonstrate that incorporating multi-band data enhances DeepDISC’s ability to detect or deblend closely spaced objects, compared to Euclid MER PF.}
\label{fig:multi_deblend}
\end{figure*}

\begin{figure*}
\centering
\includegraphics[width=0.8\textwidth]{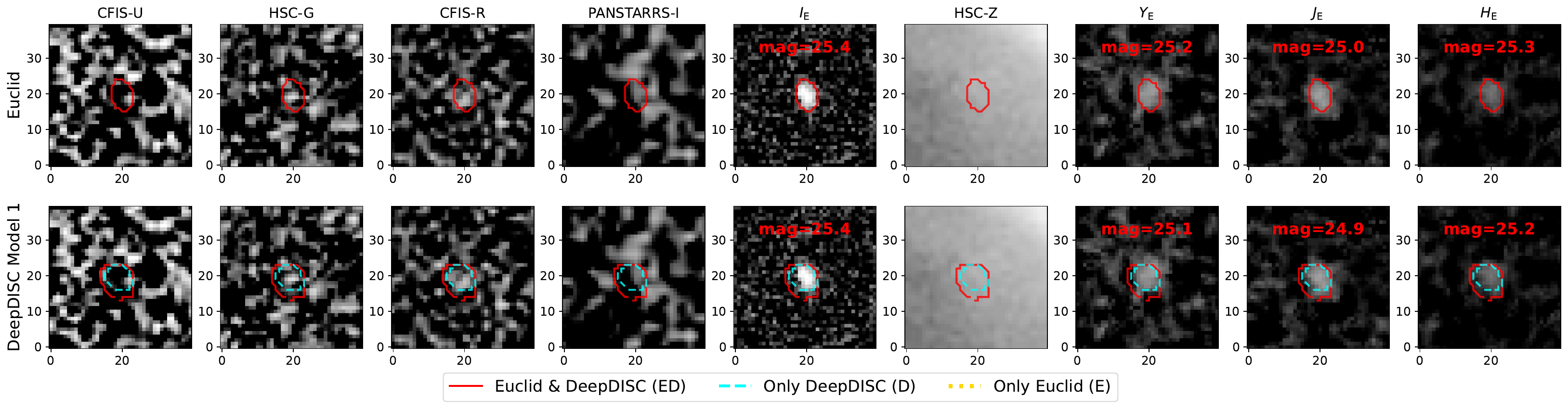}
\includegraphics[width=0.8\textwidth]{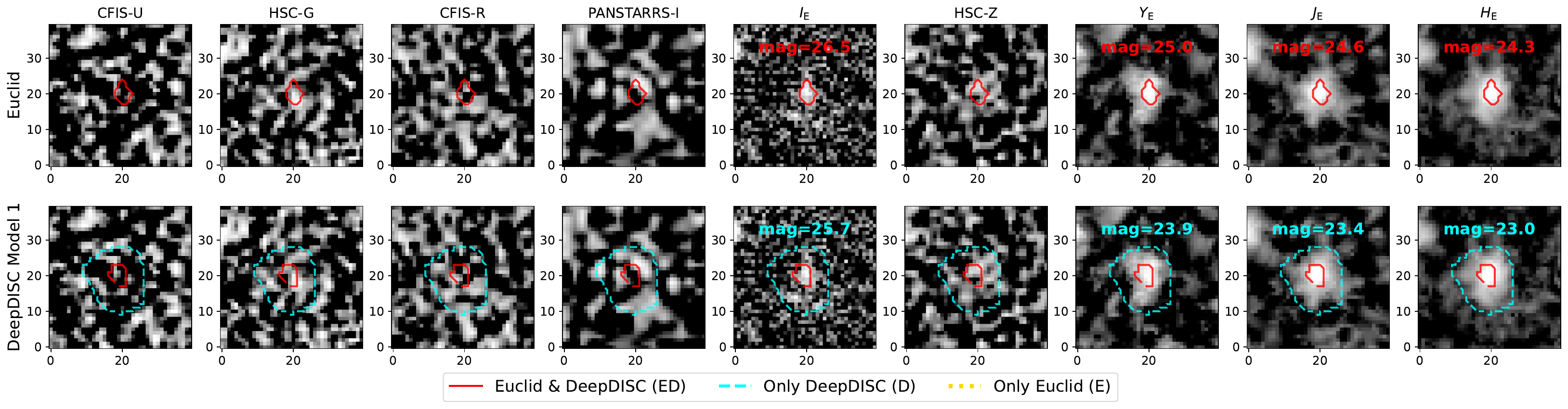}
\caption{Two examples of objects with multiple detections by DeepDISC at nearly identical positions, resulting in DeepDISC-only (D) detections. Both examples are faint in $\Ie$ band, with a magnitude larger than 24. In top panels, two detections appear similar and faint in all bands and are likely produced by noise fluctuations. In bottom panels, the objects exhibit different morphologies and magnitudes in optical and near-infrared bands, suggesting that they may correspond to two distinct objects or structures with different colors. These examples indicate that some sources faint in the $\Ie$ band but bright in other bands can still correspond to real astrophysical objects.}
\label{fig:multi_spurious}
\end{figure*}

We present several examples comparing Euclid and DeepDISC detections for discrepant cases. Figure~\ref{fig:ext} illustrates unmatched bright objects between the two methods. Figure~\ref{fig:artifact} shows that image artifacts can lead to spurious detections in both Euclid and DeepDISC. Figure~\ref{fig:deblend} provides representative deblending examples in the $\Ie$ band, while Figure~\ref{fig:multi_deblend} demonstrates that DeepDISC performs better in separating close objects with distinct colors. Finally, Figure~\ref{fig:multi_spurious} highlights a common issue in DeepDISC for faint objects ($\Ie>24$), where multiple (bogus) detections are produced at nearly the same position.

\end{appendix}

\end{document}